\documentclass[10pt,journal,twocolumn]{IEEEtran}
\usepackage{ifpdf}
 \ifpdf
 \else
 \fi
\ifCLASSINFOpdf
   \usepackage[pdftex]{graphicx}
\else
   \usepackage[dvips]{graphicx}
\fi
\usepackage[cmex10]{amsmath}

\usepackage[tight,footnotesize]{subfigure}
\usepackage{define-notations,amssymb}
\usepackage{color,algorithmic,algorithm,cite}

\newtheorem{lemma}{Lemma}

\begin{document}
\title{Weighted Sum Rate Maximization for\\
Downlink OFDMA with Subcarrier-pair based \\
Opportunistic DF Relaying}

\author{Tao Wang,~\IEEEmembership{Senior Member, IEEE},
Fran\c{c}ois Glineur, J\'er{\^o}me Louveaux and Luc Vandendorpe,~\IEEEmembership{Fellow, IEEE}
 %and Luc Vandendorpe,~\IEEEmembership{Fellow, IEEE}
\thanks{
Copyright (c) 2012 IEEE. Personal use of this material is permitted. However, permission to use this material for any other purposes must be obtained from the IEEE by sending a request to pubs-permissions@ieee.org.
Part of this paper has been presented in 2013 IEEE Wireless
Communication and Networking Conference, Shanghai, China.

T. Wang is with School of Communication \& Information Engineering,
Shanghai University, 200072 Shanghai, P. R. China.
He was with ICTEAM Institute, Universit\'e Catholique de Louvain (UCL),
1348 Louvain-la-Neuve, Belgium (Email: t.wang@ieee.org).

F. Glineur, J. Louveaux and L. Vandendorpe are with ICTEAM Institute,
UCL, 1348 Louvain-la-Neuve, Belgium (Email:\{francois.glineur, jerome.louveaux, luc.vandendorpe\}@uclouvain.be).}

\thanks{
This research is supported by The Program for Professor of
Special Appointment (Eastern Scholar) at Shanghai Institutions of Higher Learning.
It is also supported by the European Commission in the framework
of the FP7 Network of Excellence in Wireless COMmunications NEWCOM\#
(Grant agreement no. 318306), the IAP project BESTCOM, and the ARC SCOOP.

}}

\markboth{Submitted to IEEE Transactions on Signal Processing}
{WSR Maximization for...}

\maketitle

\begin{abstract}
This paper addresses a weighted sum rate (WSR) maximization problem
for downlink OFDMA aided by a decode-and-forward (DF) relay
under a total power constraint.
A novel subcarrier-pair based opportunistic DF relaying protocol is proposed.
Specifically, user message bits are transmitted in two time slots.
A subcarrier in the first slot can be paired with a subcarrier in the second slot
for the DF relay-aided transmission to a user.
In particular, the source and the relay can transmit
simultaneously to implement beamforming at the subcarrier in the second slot.
Each unpaired subcarrier in either the first or second slot is used for
the source's direct transmission to a user.
A benchmark protocol, same as the proposed one
except that the transmit beamforming is not used for the relay-aided transmission,
is also considered.
For each protocol, a polynomial-complexity algorithm is developed to
find at least an approximately optimum resource allocation (RA),
by using continuous relaxation, the dual method, and Hungarian algorithm.
Instrumental to the algorithm design is an elegant definition
of optimization variables, motivated by the idea of
{\it regarding the unpaired subcarriers as virtual subcarrier pairs
in the direct transmission mode}.
The effectiveness of the RA algorithm
and the impact of relay position and total power on
the protocols' performance are illustrated by numerical experiments.
It is shown that for each protocol,
it is more likely to pair subcarriers for relay-aided transmission
when the total power is low and the relay lies in the middle
between the source and user region.
The proposed protocol always leads to
a maximum WSR equal to or greater than that for the benchmark one,
and the performance gain of using the proposed one is significant
especially when the relay is in close proximity to the source
and the total power is low.
Theoretical analysis is presented to interpret these observations.
\end{abstract}


% For peer review papers, you can put extra information on the cover
% page as needed:
% \ifCLASSOPTIONpeerreview
% \begin{center} \bfseries EDICS Category: 3-BBND \end{center}
% \fi
%
% For peerreview papers, this IEEEtran command inserts a page break and
% creates the second title. It will be ignored for other modes.
\IEEEpeerreviewmaketitle

\section{Introduction}

Orthogonal frequency division multiple access (OFDMA)
has been widely recognized as one of the dominant wireless 
technologies for high data-rate transmission. 
One of the main reasons behind this fact is that 
spectral efficiency of the OFDM(A) systems can be improved significantly by
proper resource allocation (RA) when transmitter channel state information 
(CSI) is available \cite{Wang12TSP,Wang12ICC,Wang11TSP-1}. 
The incorporation of decode-and-forward (DF) and amplify-and-forward (AF)
relaying into OFDM(A) systems through subcarrier-pair based protocols
and associated RA have lately been under intensive investigation \cite{Tao10,Hottinen06,Herdin06,Hammerstrom07,Dang10,
WangWY08,WangWY09,Li09,Wang12EL,Vandendorpe08-1,Jerome08,Ng07,WangYing07,Li08,Haj11,ZhiWen12,ZhiWen13,
Vandendorpe08-2,Vandendorpe09-1,Vandendorpe09-2,Hsu11,WangTSP11,WangJSAC11,Boost11,Liu12,Wang13CL}.
This class of protocols share the following features.
User message bits are transmitted during two consecutive equal-duration time slots.
In the first slot, the source broadcasts OFDM symbols, so does the relay in the second slot.
The source might also emit OFDM symbols during the second slot
as will be elaborated later.
A subcarrier in the first slot can be paired with a subcarrier
in the second slot for transmitting message bits with DF/AF relaying,
referred to as the relay-aided transmission mode hereafter.

In this paper, we focus on RA for downlink OFDMA
with subcarrier-pair based DF relaying 
(there also exist works on RA for OFDMA systems using bidirectional 
relaying \cite{Tao10}). 
The subcarrier-pair based AF relaying has been studied
in \cite{Hottinen06,Herdin06,Hammerstrom07,Dang10}.
Note that the subcarrier-by-subcarrier based pairing
may not be sufficient for DF relaying, since the information from a set of subcarriers
in the first time slot can be decoded and re-encoded jointly
and then forwarded through a different set of subcarriers
in the second time slot \cite{Dang10,Wang12EL}.
Nevertheless, the subcarrier-pair based DF relaying
has attracted much research interest due to simplicity or practical reasons
\cite{Wang12EL,WangWY08,WangWY09,Li09,
Vandendorpe08-1,Jerome08,Ng07,WangYing07,Li08,Haj11,ZhiWen12,ZhiWen13,
Vandendorpe08-2,Vandendorpe09-1,Vandendorpe09-2,Hsu11,WangTSP11,WangJSAC11,Boost11,Liu12,Wang13CL}.

\begin{figure}%[h]
  \centering
  \subfigure[when the S-D link is unavailable \cite{Wang12EL,WangWY08,WangWY09,Li09}.]{
     \includegraphics[width=3in]{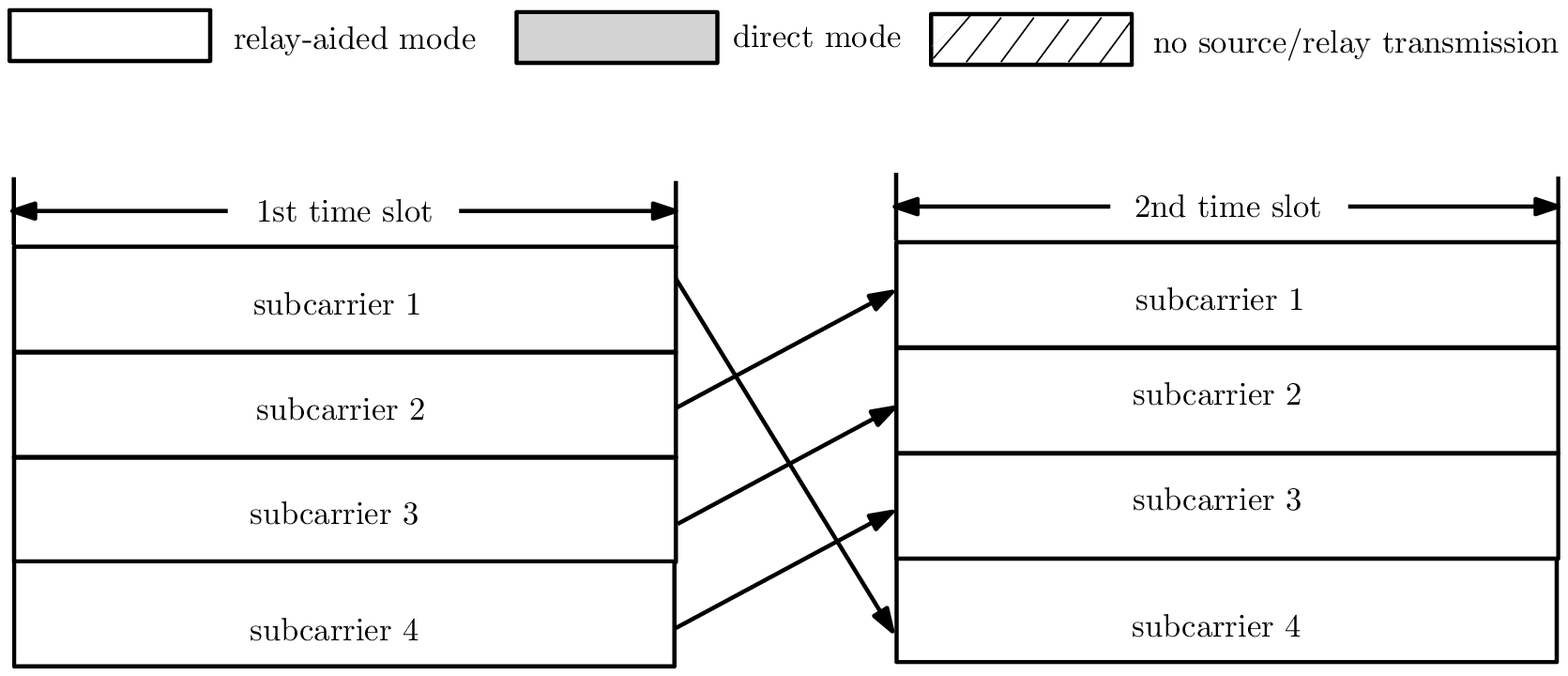}}
  \subfigure[when the S-D link is available but the source does not transmit in the second slot \cite{Vandendorpe08-1,Jerome08,Ng07,WangYing07,Li08,Haj11,ZhiWen12,ZhiWen13}.]{
     \includegraphics[width=3in]{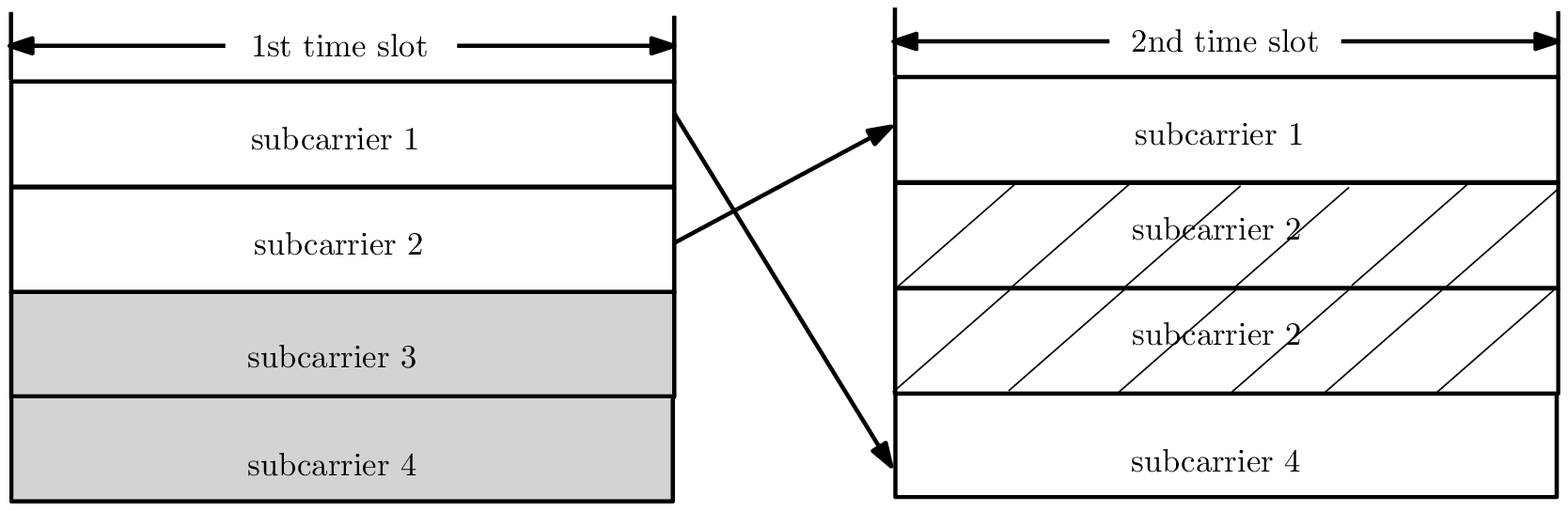}}
  \subfigure[when the S-D link is available and the source transmits in the second slot \cite{Vandendorpe08-2,Vandendorpe09-1,Vandendorpe09-2,Hsu11,WangTSP11,WangJSAC11,Boost11,Liu12,Wang13CL}.]{
     \includegraphics[width=3in]{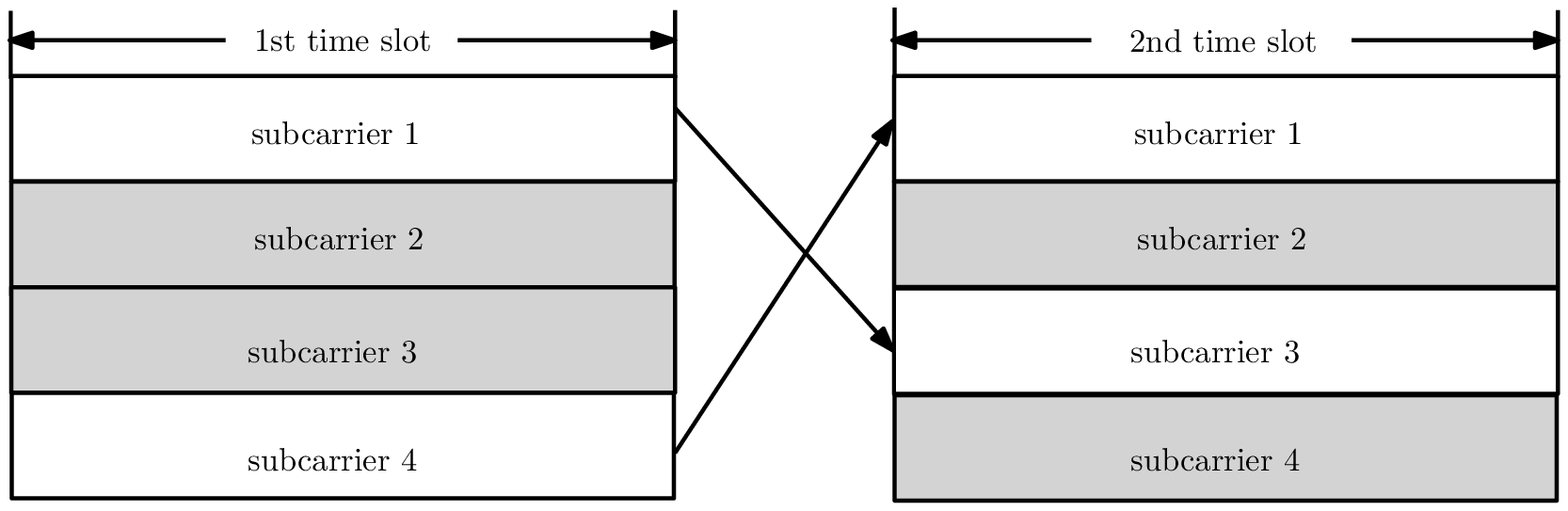}}
  \caption{Illustration of the subcarrier-pair based DF relaying protocols for OFDM(A)-based systems, where every arrow indicates that the two associated subcarriers
  are paired for the relay-aided transmission.}  \label{fig:protocols}
\end{figure}

When the source-to-destination (S-D) link is unavailable (i.e., the destination lies outside
the source's radio coverage), RA problems for OFDM systems using subcarrier-pair based DF protocols
have been addressed in \cite{WangWY08,WangWY09,Li09,Wang12EL}.
In these works, every subcarrier in the first time slot is paired with
a subcarrier in the second time slot for the relay-aided transmission, as illustrated in Fig. \ref{fig:protocols}.a.
To maximize sum rate under a total power constraint,
ordered subcarrier pairing has been proven to be the optimum, i.e.,
the strongest source-to-relay subcarrier should be paired with
the strongest relay-to-destination subcarrier, and so on.

The works in \cite{Vandendorpe08-1,Jerome08,Ng07,WangYing07,Li08,Haj11,ZhiWen12,ZhiWen13,
Vandendorpe08-2,Vandendorpe09-1,Vandendorpe09-2,Hsu11,WangTSP11,WangJSAC11,Boost11,Liu12,Wang13CL}
have considered the case where the S-D link is available.
When only the relay emits OFDM symbols in the second time slot,
opportunistic relaying (sometimes termed as selection relaying) was studied in
\cite{Vandendorpe08-1,Jerome08,Ng07,WangYing07,Li08,Haj11,ZhiWen12,ZhiWen13}.
Specifically, a subcarrier in the first time slot can either be
paired with a subcarrier in the second slot for the relay-aided transmission,
or used directly for the S-D transmission without the relay's assistance,
referred to as the direct transmission mode hereafter.
It is very important to note that when some subcarriers
in the first slot are used in the direct transmission mode,
some subcarriers in the second slot will not be used
as illustrated in Fig. \ref{fig:protocols}.b,
which leads to a waste of precious spectrum resource.

To address the above issue, improved protocols which allow the source
to emit OFDM symbols in the second slot were proposed and studied in
\cite{Vandendorpe08-2,Vandendorpe09-1,Vandendorpe09-2,WangTSP11,WangJSAC11,Hsu11,Boost11,Liu12,Wang13CL}.
The improved protocols are the same as those considered in \cite{Vandendorpe08-1,Jerome08,Ng07,WangYing07,Li08,Haj11},
except that the source can also make direct S-D transmission
at every unpaired subcarrier in the second slot, as illustrated in Fig. \ref{fig:protocols}.c.
Note that the improved protocols do {\it not} really
improve the way that DF relaying is implemented over a subcarrier pair,
but rather let the source utilize the unpaired subcarriers in the second slot
for direct transmission to avoid the waste of spectrum resource.
In \cite{Hsu11,Boost11,Wang13CL}, the subcarrier pairing and power allocation
are jointly optimized for point-to-point OFDM systems.
As for OFDMA systems, RA problems considering the joint optimization
of power allocation, subcarrier assignment to users
and selection of multiple relays for transmit beamforming in the second slot 
are addressed in \cite{WangTSP11,WangJSAC11}.
In these works, a priori and CSI-independent subcarrier pairing
is considered, i.e., a subcarrier
in the first slot is always paired with the same subcarrier in the second slot
if the relay-aided mode is used.
The optimization of subcarrier pairing and assignment to users
is addressed in \cite{Liu12} with a graph based approach.
It is a complicated RA problem to jointly optimize subcarrier pairing
and mode selection with power allocation and subcarrier assignment to users.

Compared with the above existing works, this paper makes the following contributions:

\begin{itemize}
\item
A novel subcarrier-pair based opportunistic DF protocol is proposed
for downlink OFDMA aided by a DF relay.
This protocol further makes improvement over those previously studied in the literature
\cite{Vandendorpe08-2,Vandendorpe09-1,Vandendorpe09-2,Hsu11,WangTSP11,WangJSAC11,Boost11,Liu12,Wang13CL},
by allowing the source and the relay to implement transmit beamforming
at a subcarrier in the second time slot for the relay-aided transmission.
Note that the protocols studied in \cite{WangTSP11,WangJSAC11}
considered the selection of multiple DF relays ({\it excluding the source}) 
for transmit beamforming in the second slot, 
while the proposed protocol considers the joint source-relay transmit beamforming. 
A benchmark protocol, which is the same as the proposed one
except for the relay-aided transmission mode, is also considered.
Note that the proposed protocol {truly} improves the implementation of
DF relaying over a subcarrier pair with transmit beamforming,
which is not the case for the benchmark protocol.

\item
The weighted sum rate (WSR) maximized RA problem
is addressed for both the proposed and benchmark protocols
under a total power constraint for the whole system.
First, it is shown that the proposed protocol leads to a maximum WSR
not smaller than that for the benchmark one.
Then, an algorithm is developed for each protocol
to find at least an approximately optimum RA with a WSR
very close to the maximum WSR.
Instrumental to the elegance of the RA algorithm is a definition
of appropriate indicator variables, making it possible to
cast a subproblem related to the joint optimization
of transmission-mode selection, subcarrier pairing and assignment to users
into an standard assignment problem that can be solved efficiently
by Hungarian algorithm.
\end{itemize}

The rest of this paper is organized as follows.
In the next section, the system and transmission protocols are described.
The theoretical analysis is made to compare
the maximum WSRs of the two protocols in Section \ref{sec:insight}.
After that, the RA algorithm is developed in Section \ref{sec:RA}. 
Numerical experiments are shown to illustrate the effectiveness 
of the RA algorithm and study the impact of relay position and total power 
on the protocols' performance in Section \ref{sec:numexp}. 
Finally, some conclusions are drawn.

Notations: A letter in bold, e.g. $\bf x$, represents a set.
$\Rate{x} = \frac{1}{2}\log_2(1 + x)$.

\section{Protocols and WSR maximization problem}

\subsection{The transmission system and protocols}

Consider the downlink OFDMA transmission from a source to $U$ users
(user $u=1,\dots,U$) aided by a DF relay.
The source, relay and every user are each equipped with a single antenna,
and the channel between every two of them is frequency selective.
The source and the relay are synchronized so that they can simultaneously emit
OFDM symbols using $K$ subcarriers and with sufficiently long cyclic prefix
to eliminate inter-symbol interference.

The novel transmission protocol is half-duplex,
i.e., user message bits are transmitted in two consecutive
equal-duration time slots, during which all channels are assumed to keep unchanged.
During the first slot, only the source broadcasts $N$ OFDM symbols.
Both the relay and all users receive these symbols.
After proper processing explained later, the source and relay simultaneously
broadcast $N$ OFDM symbols, and the users receive them during the second slot.

Due to the OFDMA, each subcarrier is dedicated to transmitting
a single user's message exclusively.
A subcarrier in the first slot can be paired with a subcarrier in the second slot
for the relay-aided mode transmission to a user.
Each unpaired subcarrier in either the first or second slot is used by the source
for the direct mode transmission to a user.

To simplify description, we use subcarriers $k$ and $l$ to denote
the $k$th and $l$th subcarriers used during the first and second slots, respectively ($k,l=1,\cdots,K$).
We define the source transmission powers for subcarrier $k$
in the first slot and subcarrier $l$ in the second slot as $\Pskone$ and $\Psltwo$, respectively.
The relay transmission power for subcarrier $l$ is $\Prl$.
The complex amplitude gains at subcarrier $k$
for the source-to-relay, source-to-$u$ and relay-to-$u$ channels
are $\hsrk$, $\hsuk$ and $\hruk$, respectively.
The two transmission modes for the novel protocol are elaborated as follows:

\medskip
\subsubsection{The relay-aided transmission mode}

\begin{figure}
  \centering
  \subfigure[]
  {\includegraphics[width=1.7in]{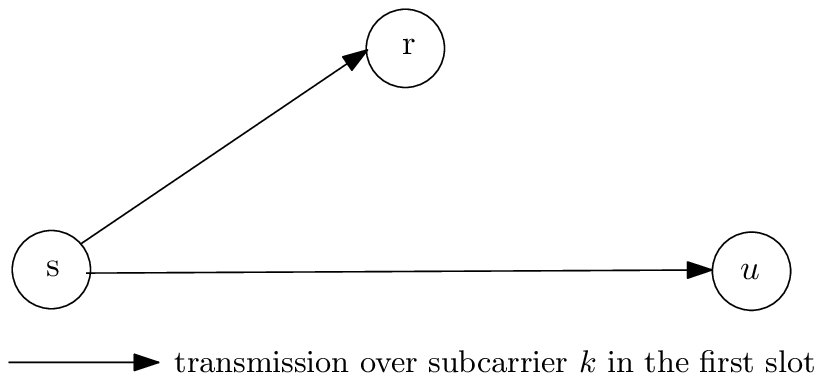}}
  \subfigure[]
  {\includegraphics[width=1.7in]{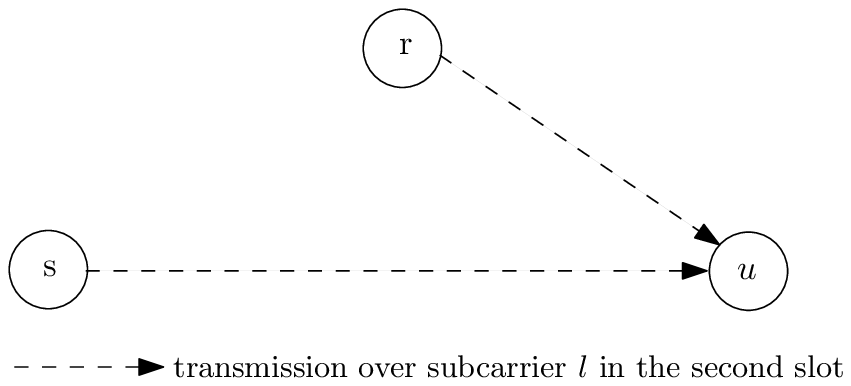}}
  \caption{The relay-aided transmission mode over the subcarrier pair $(k,l)$ to user $u$.}\label{fig:relay-mode}
\end{figure}

Suppose subcarrier $k$ is paired with subcarrier $l$ for the relay-aided mode transmission
to user $u$. In such a case, we refer to the two subcarriers collectively as the subcarrier pair $(k,l)$.
A block of message bits are first encoded into a code word of complex symbols
$\{\Code(n)|n=1,\cdots,N\}$ with $E(|\Code(n)|^2)=1$, $\forall\;n$.
In the first slot, the source broadcasts the codeword over subcarrier $k$
as illustrated in Figure \ref{fig:relay-mode}.a.
At the relay and user $u$, the $n$th baseband signals received through subcarrier $k$ are
\begin{align}
\Yrk(n) = \sqrt{\Pskone} \hsrk\Code(n) + \Nrk(n), n = 1,\cdots,N,
\end{align}
and
\begin{align}
\Yukone(n) = \sqrt{\Pskone}\hsuk\Code(n) + \Nukone(n), n = 1,\cdots,N,
\end{align}
respectively, where $\Nrk(n)$ and $\Nukone(n)$ are both additive white Gaussian noise (AWGN)
with power $\sigma^2$. The signal-to-noise ratio (SNR) at the relay is
$\Pskone\Gsrk$ where $\Gsrk=\frac{|\hsrk|^2}{\sigma^2}$.
At the end of the first time slot, the relay decodes the message bits
from $\{\Yrk(n)|n = 1,\cdots,N\}$ and
then reencodes those bits into the same codeword as the source did.

In the second time slot, the source and relay broadcast the codewords
$\{\Code(n)e^{-j\angle\hsul}|\forall\;n\}$ and $\{\Code(n)e^{-j\angle\hrul}|\forall\;n\}$
through subcarrier $l$, respectively, where $\angle\hsul$ and $\angle\hrul$ represent
the phase of $\hsul$ and $\hrul$, respectively.
This means that the source and relay implement transmit beamforming
to emit the codeword through subcarrier $l$ as illustrated in Figure \ref{fig:relay-mode}.b.
Note that the source and relay need to know
the phase of $\hsul$ and $\hrul$, respectively.
At user $u$, the $n$th baseband signal received through subcarrier $l$ is
\begin{align}
\Yultwo(n) = \big(\sqrt{\Psltwo}|\hsul|+\sqrt{\Prl}|\hrul|\big)\Code(n) + \Nultwo(n),
\end{align}
where $\Nultwo(n)$ is the AWGN with power $\sigma^2$.

Finally, user $u$ decodes the message bits from all signals received during the two slots.
These signals can be grouped into $N$ vectors, the $n$th of which is
\begin{align}
\Yn &= \left[\begin{array}{c}
                \Yukone(n) \\
                \Yultwo(n) \end{array}\right]      \label{eq:eqv-channel}  \\
    &= \left[\begin{array}{c}
           \sqrt{\Pskone}\hsuk  \\
           \sqrt{\Psltwo}|\hsul|+\sqrt{\Prl}|\hrul|  \end{array}\right] \Code(n) + \Nn,   \nonumber
\end{align}
where $\Nn = [\Nukone(n),\Nultwo(n)]^T$.
Note that the transmission in effect makes $N$ uses of a discrete memoryless single-input-two-output
channel specified by \eqref{eq:eqv-channel}, with the $n$th input and output being $\Code(n)$ and $\Yn$, respectively.
To achieve the maximum reliable transmission rate,
maximum ratio combining should be used \cite{Fund-WCOM},
i.e., user $u$ first turns every $\Yn$ into a decision variable
\begin{align}
&\MRCn = (\sqrt{\Pskone}\hsuk)^*\Yukone(n) +  \nonumber\\
&\hspace{2cm}         \big(\sqrt{\Psltwo}|\hsul|+\sqrt{\Prl}|\hrul|\big)^*\Yultwo(n),
\end{align}
and then decodes the message from $\{\MRCn| \forall\;n\}$.
It can readily be derived that the SNR for this decoding is
\begin{align}
&\SNRklu(\Pskone,\Psltwo,\Prl) = \Gsuk\Pskone +   \nonumber\\
&\hspace{2cm}         \big(\sqrt{\Gsul\Psltwo}+\sqrt{\Grul\Prl}\big)^2,
\end{align}
where $\Gsuk=\frac{|\hsuk|^2}{\sigma^2}$ and $\Grul=\frac{|\hrul|^2}{\sigma^2}$.

To ensure both the relay and user $u$ can reliably decode the message bits,
the maximum number of message bits that can be transmitted is $2N\Rate{\Gsrk\Pskone}$
and $2N\Rate{\SNRklu(\Pskone,\Psltwo,\Prl)}$, respectively.
This means that the maximum transmission rate over the subcarrier pair $(k,l)$
in the relay-aided mode to user $u$ is equal to
$\Rate{\min\{\Gsrk\Pskone, \SNRklu(\Pskone,\Psltwo,\Prl)\}}$ bits/OFDM-symbol
(bpos)\footnote{Recall that $2N$ OFDM symbols are used in total during the two time slots.}.

\subsubsection{The direct transmission mode}

Suppose subcarrier $k$ (respectively, subcarrier $l$) is unpaired with any subcarrier
in the second (respectively, first) slot, and is used for direct mode transmission to user $u$.
The source first encodes message bits into a codeword of $N$ symbols,
which are then broadcast through subcarrier $k$ (respectively, subcarrier $l$.
In such a case, the relay keeps silent at subcarrier $l$ in the second slot, i.e. $\Prl=0$.).
User $u$ decodes the message bits from the signals received through subcarrier $k$
(respectively, subcarrier $l$).
The maximum rate through subcarrier $k$ (respectively, subcarrier $l$) in the direct transmission mode
is $\Rate{\Pskone\Gsuk}$ (respectively, $\Rate{\Psltwo\Gsul}$) bpos.

\medskip
A benchmark protocol is also considered.
This protocol is the same as the novel protocol
except for the relay-aided transmission mode.
Specifically, the relay-aided mode is the same as that
widely studied in the literature \cite{Vandendorpe08-1,Jerome08,Ng07,WangYing07,Li08,Haj11,
Vandendorpe08-2,Vandendorpe09-1,Vandendorpe09-2,Hsu11,WangTSP11,WangJSAC11,Boost11},
i.e., the source does not transmit at subcarrier $l$ during the second slot,
if subcarriers $k$ and $l$ are paired for the relay-aided transmission to user $u$.
In such a case, the maximum rate for the relay-aided transmission
over that subcarrier pair to user $u$ is equal to
$\Rate{\min\{\Gsrk\Pskone,\Gsuk\Pskone + \Grul\Prl\}}$ bpos.
It is important to note that, the benchmark protocol
is a special case of the novel protocol, since it is equivalent
to the novel protocol with the constraint that $\Psltwo=0$
if subcarrier $l$ is paired with a subcarrier in the first slot
for the relay-aided mode transmission.

\subsection{The WSR maximization problem}

We assume there exists a central controller which knows precisely the CSI
$\{\Gsrk,\Gsuk,\Gruk|\forall\;k\}$. Before the data transmission,
the controller needs to find the optimum subcarrier and power assignment,
i.e., which subcarriers should be paired for the relay-aided mode
and which should be in the direct mode,
how these subcarriers should be assigned to the users,
as well as the source/relay power allocation to maximize the WSR of all users
for the adopted transmission protocol (which can be either the novel or benchmark protocol),
when the total power consumption is not higher than a prescribed value $\Ptot$.
Then, the controller can inform the source and the relay
about the optimum subcarrier and power assignment to be adopted for data transmission.

\section{Theoretical analysis}\label{sec:insight}

It can be shown that the proposed protocol leads to
a maximum WSR greater than or equal to that for the benchmark protocol.
To this end, suppose the optimum subcarrier assignment and power allocation
has been found for the benchmark protocol.
By using the proposed protocol with the same subcarrier assignment and power allocation,
the same WSR can be achieved.
Obviously, the maximum WSR for the proposed protocol is greater than or equal to
that WSR, namely the maximum WSR for the benchmark protocol.

In Section \ref{sec:pair-ratemax}, we assume
subcarriers $k$ and $l$ are paired for the relay-aided mode transmission to user $u$,
and a sum power $P$ is used for this pair.
We focus on computing the maximum rate
and optimum power allocation of this pair for both protocols.
Using these results, theoretical analysis will be made in Section \ref{sec:comparison}
to show when the maximum WSR for the proposed protocol is strictly greater
than that for the benchmark one,
and the RA algorithm will be developed in Section \ref{sec:RA}.
Moreover, this analysis plays an important role to interpret the numerical
experiments shown in Section \ref{sec:numexp} to illustrate
the impact of the relay's position on the benefit of using the proposed protocol.

\subsection{Rate maximization for the pair in the relay-aided mode}\label{sec:pair-ratemax}

\subsubsection{Analysis for the proposed protocol}
To facilitate derivation, define $\diffGuk = \Gsrk - \Gsuk$ and $\sumGul = \Gsul + \Grul$.
To maximize the rate, the optimum $\Pskone$, $\Psltwo$ and $\Prl$ are the optimum solution for
\begin{align}
\max_{\Pskone,\Psltwo,\Prl}  &\hspace{0.2cm}  \min\{\Gsrk\Pskone, \SNRklu(\Pskone,\Psltwo,\Prl)\} \nonumber  \\
{\rm s.t.}                   &\hspace{0.2cm}  \Pskone+\Psltwo+\Prl = P,    \label{prob:novel-kl}\\
                             &\hspace{0.2cm}  \Pskone\geq0,\Psltwo\geq0,\Prl\geq0.      \nonumber
\end{align}

\begin{figure}[h]
\centering  \includegraphics[width=3.5in]{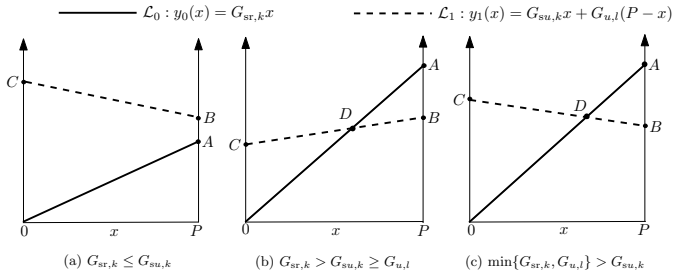}
\caption{Illustration of the two lines $\mathcal{L}_0$ and $\mathcal{L}_1$ in three different cases.} \label{fig:novel-kl}
\end{figure}

By using the Cauchy-Schwartz inequality, it can be shown that
\begin{align}
\SNRklu(\Pskone,\Psltwo,\Prl) \leq \Gsuk\Pskone + \sumGul P_2,  \label{eq:inequality}
\end{align}
where $P_2 = \Psltwo + \Prl$ and the inequality is tight when
$\Psltwo = \frac{\Gsul}{\sumGul}P_2$ and $\Prl = \frac{\Grul}{\sumGul}P_2$.
Now, the optimum solution for \eqref{prob:novel-kl} can be found by first solving
\begin{align}
\max_{\Pskone,P_2}  &\hspace{0.2cm}  \min\{\Gsrk\Pskone, \Gsuk\Pskone + \sumGul P_2\}  \label{prob:novel-kl-2} \\
{\rm s.t.}          &\hspace{0.2cm}  \Pskone + P_2 = P, \Pskone\geq0,P_2\geq0     \nonumber
\end{align}
for the optimum $\Pskone$ and $P_2$, and then using that $P_2$ to compute
the optimum $\Psltwo$ and $\Prl$ according to the formulas that tighten the inequality \eqref{eq:inequality}.
Problem \eqref{prob:novel-kl-2} can be solved intuitively as follows.
First, the two lines
\begin{align}
\mathcal{L}_0 &= \{(x, y_0(x))| x\in[0,P], y_0(x) = \Gsrk x\}     \nonumber \\
\mathcal{L}_1 &= \{(x, y_1(x))| x\in[0,P], y_1(x) = \Gsuk x + \sumGul(P-x)\}  \nonumber
\end{align}
can be plot over the two-dimensional plane of coordinates $(x,y)$ in Fig. \ref{fig:novel-kl}.
It can be seen that three different cases are possible,
each corresponding to a specific orientation of the two lines.
The coordinates of points $A$, $B$, $C$ and $D$ in the figure are shown in Table \ref{tab:ABCD}.
The optimum $\Pskone$ and objective value for \eqref{prob:novel-kl-2} (which are also for \eqref{prob:novel-kl})
are equal to the $x$ and $y$ coordinates of the points $A$, $B$ and $D$ for the three cases in Fig. \ref{fig:novel-kl}, respectively.
From this fact, it can easily be seen that the optimum $\Pskone$, $\Psltwo$ and $\Prl$
for \eqref{prob:novel-kl} are
\begin{align}
\Pskone = \left\{\begin{array}{ll}
               \frac{\sumGul}{\diffGuk + \sumGul}P & {\rm if\;} \min\{\Gsrk,\sumGul\}> \Gsuk,  \\
               P                                 & {\rm if\;}   \min\{\Gsrk,\sumGul\}\leq \Gsuk,
           \end{array}\right. \nonumber
\end{align}
\begin{align}
\Psltwo = \left\{\begin{array}{ll}
               \frac{\Gsul}{\sumGul}\frac{\diffGuk}{(\diffGuk + \sumGul)}P & {\rm if\;} \min\{\Gsrk,\sumGul\}> \Gsuk,  \\
               0                                              & {\rm if\;} \min\{\Gsrk,\sumGul\}\leq \Gsuk,
           \end{array}\right. \nonumber
\end{align}
and
\begin{align}
\Prl = \left\{\begin{array}{ll}
               \frac{\Grul}{\sumGul}\frac{\diffGuk}{(\diffGuk + \sumGul)}P & {\rm if\;}\min\{\Gsrk,\sumGul\}> \Gsuk,  \\
               0     & {\rm if\;}\min\{\Gsrk,\sumGul\}\leq \Gsuk.
           \end{array}\right. \nonumber
\end{align}

The maximum rate associated with the above optimum solution is equal to $\Rate{\newGklu P}$ with
\begin{align}
\newGklu = \left\{\begin{array}{ll}
               \frac{\Gsrk\sumGul}{\diffGuk + \sumGul}    & {\rm if\;}\min\{\Gsrk,\sumGul\}> \Gsuk,  \\
               \min\{\Gsrk, \Gsuk\}                       & {\rm if\;}\min\{\Gsrk,\sumGul\}\leq \Gsuk .
           \end{array}\right. \label{eq:newGklu}
\end{align}

\begin{table}[h]
  \centering
  \caption{Coordinates of A, B, C and D in Figure \ref{fig:novel-kl}.}\label{tab:ABCD}
  \begin{tabular}{c||c|c|c|c}
    \hline
             &   $A$       & $B$         & $C$           & $D$          \\\hline\hline
     $x$     &   $P$       & $P$         & $0$           & $\frac{\sumGul}{\diffGuk + \sumGul}P$  \\\hline
     $y$     &   $\Gsrk P$ & $\Gsuk P$   & $\sumGul P$   & $\frac{\Gsrk\sumGul}{\diffGuk + \sumGul}P$   \\\hline
  \end{tabular}
\end{table}

\subsubsection{Analysis for the benchmark protocol}
In this case, $\Psltwo=0$ and the optimum $\Pskone$ and $\Prl$
for maximizing the rate are the optimum solution for
\begin{align}
\max_{\Pskone,\Prl}  &\hspace{0.2cm}  \min\{\Pskone\Gsrk,\Pskone\Gsuk + \Prl\Grul\}    \label{prob:ben-kl}\\
{\rm s.t.}           &\hspace{0.2cm}  \Pskone + \Prl = P,\Pskone\geq0,\Prl\geq0,    \nonumber
\end{align}
which can also be solved by the intuitive method as described above.
It can be shown that the optimum $\Pskone$ and $\Prl$ are
\begin{align}
\Pskone = \left\{\begin{array}{ll}
               \frac{\Grul}{\diffGuk + \Grul}P   & {\rm if\;}\min\{\Gsrk,\Grul\}> \Gsuk,  \\
               P                                 & {\rm if\;}\min\{\Gsrk,\Grul\}\leq \Gsuk,
           \end{array}\right. \nonumber
\end{align}
and
\begin{align}
\Prl = \left\{\begin{array}{ll}
               \frac{\diffGuk}{\diffGuk + \Grul}P  & {\rm if\;}\min\{\Gsrk,\Grul\}> \Gsuk,  \\
               0                                   & {\rm if\;}\min\{\Gsrk,\Grul\}\leq \Gsuk,
           \end{array}\right. \nonumber
\end{align}
and the maximum rate associated with the above optimum solution is equal to $\Rate{\oldGklu P}$ with
\begin{align}
\oldGklu = \left\{\begin{array}{ll}
               \frac{\Gsrk\Grul}{\diffGuk + \Grul}  & {\rm if\;}\min\{\Gsrk,\Grul\}> \Gsuk,  \\
               \min\{\Gsrk,\Gsuk\}                  & {\rm if\;}\min\{\Gsrk,\Grul\}\leq \Gsuk.
           \end{array}\right. \label{eq:oldGklu}
\end{align}

\subsection{Comparison of the two protocols}\label{sec:comparison}

To compare the maximum WSR for the two protocols,
it is necessary to first compare $\newGklu$ and $\oldGklu$.
When $\Gsuk\geq \min\{\Gsrk,\Grul\}$, $\oldGklu = \min\{\Gsrk,\Gsuk\}$.
If $\min\{\Gsrk,\sumGul\}\leq \Gsuk$, $\newGklu=\min\{\Gsrk,\Gsuk\} = \oldGklu$ follows.
If $\min\{\Gsrk,\sumGul\}> \Gsuk$, it can be seen that
$\newGklu P$ and $\Gsuk P$ correspond to the $y$-coordinates of points $D$ and $B$
in Figure \ref{fig:novel-kl}.c, respectively,
and therefore $\newGklu > \Gsuk$ since $D$ is higher than $B$.
This means that $\newGklu\geq \oldGklu$ always holds when $\Gsuk\geq \min\{\Gsrk,\Grul\}$.

\begin{figure}[h]
\centering  \includegraphics[width=3.5in]{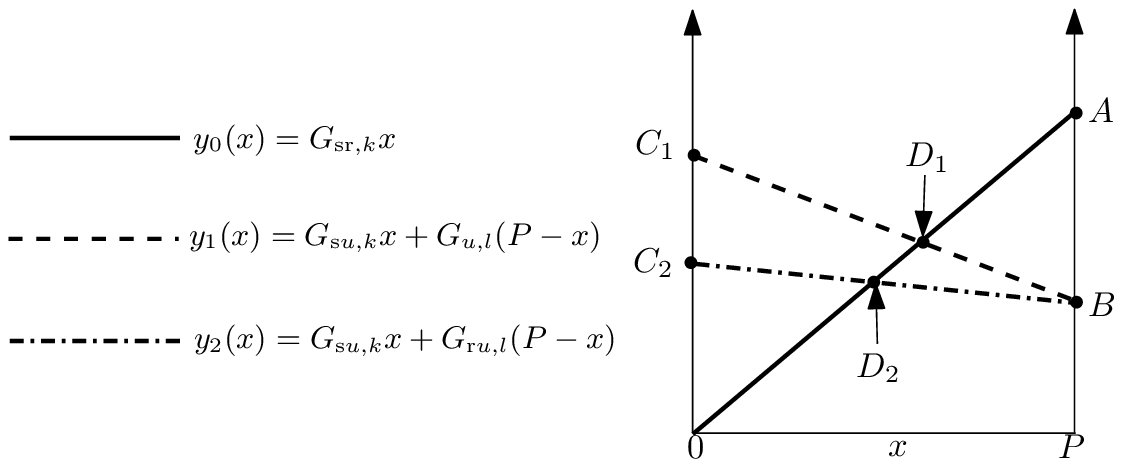}
\caption{Illustration of $\oldGklu$ and $\newGklu$ when $\min\{\Gsrk,\Grul\}> \Gsuk$.}\label{fig:compare-kl}
\end{figure}

When $\min\{\Gsrk,\Grul\}> \Gsuk$, $\newGklu$ and $\oldGklu$ can be compared
through a visualization method as follows. Specifically, we plot the lines
$\mathcal{L}_0$, $\mathcal{L}_1$ and
\begin{align}
\mathcal{L}_2 = \{(x, y_2(x))| & x\in[0,P], \\
                               & y_2(x) = \Gsuk x + \Grul (P-x) \}  \nonumber
\end{align}
in Fig. \ref{fig:compare-kl}.
The coordinates of points $A$ and $B$ are the same as given in Tab. \ref{tab:ABCD},
and those of points $C_1$, $C_2$, $D_1$ and $D_2$ are given in Tab. \ref{tab:CD}.
Most interestingly, $\newGklu P$ and $\oldGklu P$ are equal to the $y$-coordinate of
$D_1$ and $D_2$, respectively, and $C_1$ is above $C_2$ since $\sumGul\geq \Grul$.
In particular, the following points should be noted:
\begin{itemize}
\item
$\newGklu>\oldGklu$ holds because $D_1$ is above $D_2$.

\item
when $\Gsrk$ increases (meaning that point $A$ is elevated),
$\newGklu-\oldGklu$ increases (since the difference of
the $y$-coordinate of points $D_1$ and $D_2$ is increased).

\item
when $\Grul$ increases (meaning that points $C_1$ and $C_2$ are both elevated),
$\newGklu-\oldGklu$ reduces, because
\begin{align}
\newGklu - \oldGklu = \frac{\diffGuk\Gsul\Gsrk}{(\diffGuk+\Gsul+\Grul)(\diffGuk+\Grul)},  \nonumber
\end{align}
is a decreasing function of $\Grul$.
\end{itemize}

\begin{table}[h]
  \centering
  \caption{Coordinates of $C_1$, $C_2$, $D_1$ and $D_2$ in Fig. \ref{fig:compare-kl}.}\label{tab:CD}
  \begin{tabular}{c||c|c|c|c}
    \hline
         &  $C_1$      & $C_2$       & $D_1$                               & $D_2$          \\\hline\hline
    $x$  &  $0$        & $0$         & $\frac{\sumGul}{\diffGuk+\sumGul}P$ & $\frac{\Grul}{\diffGuk+\Grul}P$ \\\hline
    $y$  & $\sumGul P$ & $\Grul P$   & $\newGklu P$                        & $\oldGklu P$  \\\hline
  \end{tabular}
\end{table}

The above analysis indicates that $\newGklu\geq \oldGklu$ always holds,
and $\newGklu-\oldGklu$ increases when either $\Gsrk$ increases or $\Grul$ reduces,
if $\min\{\Gsrk,\Grul\}> \Gsuk$.

Using the above results, we now show that
the proposed protocol leads to a strictly higher maximum WSR than the benchmark protocol,
if there exist at least two subcarriers that must be paired
for the relay-aided transmission for the benchmark protocol to maximize the WSR.
To this end, collect the subcarrier pairs that must be used
by the benchmark protocol to maximize the WSR in the set $\Setsp$,
and $\forall\;(k,l)\in\Setsp$, denote $u_{kl}$ and $\oldPklukl$
as the user which should use this subcarrier pair and the sum power that should be assigned to this pair.
The rate contributed by this pair must be equal to
$\Rate{\oldGklukl \oldPklukl}$ as shown earlier.
In such a case, $\min\{\Gsrk,\Grukl\}> \Gsukl$ must be satisfied,
because otherwise simply using subcarriers $k$ and $l$ separately in the direct mode
can lead to a higher sum rate.
Suppose the proposed protocol is now used with a suboptimum RA
which adopts the same subcarrier assignment as the optimum RA for the benchmark protocol.
For every subcarrier in the direct mode, this RA uses the same source power allocation
as the optimum value for the benchmark protocol,
and $\forall\;(k,l)\in\Setsp$, this RA uses $\oldPklukl$
as the sum power for the subcarrier pair $(k,l)$.
The maximum rate for this subcarrier pair is equal to $\Rate{\newGklukl\oldPklukl}$.
Since $\min\{\Gsrk,\Grukl\}> \Gsukl$ holds,
$\newGklukl> \oldGklukl$ follows from earlier analysis,
and therefore $\Rate{\newGklukl \oldPklukl} > \Rate{\oldGklukl \oldPklukl}$ must hold.
This means that the proposed protocol has a strictly higher maximum WSR
than the benchmark protocol.

\section{RA Algorithm design}\label{sec:RA}

\subsection{Formulation of the RA problem}

To formulate the WSR maximization problem for
the adopted protocol (which can be either the proposed or benchmark protocol),
we define
\begin{align}
\Gklu = \left\{\begin{array}{ll}
                   \newGklu & {\rm if\;the\;proposed\;protocol\;is\;adopted},  \\
                   \oldGklu & {\rm if\;the\;benchmark\;protocol\;is\;adopted}.
           \end{array}\right. \nonumber
\end{align}

For any configuration of transmission-mode selection, 
subcarrier pairing and assignment to users used by the adopted protocol,
suppose $m$ subcarrier pairs are assigned to the relay-aided transmission,
then it is always possible to one-to-one associate the unpaired subcarriers
in the two slots to form $K-m$ {\it virtual subcarrier pairs}, 
each allocated to possibly two different users for direct transmission separately. 
Motivated by this observation, the RA problem is formulated by defining
the following variables:
\begin{itemize}
\item
$\tklu\in \{0,1\}$ for any combination of $k,l,u$. $\tklu=1$ indicates that
subcarrier $k$ is paired with subcarrier $l$ for the relay-aided transmission to user $u$.

\item
$\Pklu\geq 0$ for any combination of $k,l,u$.
When $\tklu=1$, $\Pklu$ is used as the total power for the subcarrier pair $(k,l)$.

\item
$\tklab\in \{0,1\}$ for any combination of $k,l$ and $a,b\in\Uset$.
$\tklab=1$ indicates that subcarrier $k$ is assigned in the direct transmission mode to user $a$
during the first slot, and so is subcarrier $l$ to user $b$ during the second slot.

\item
$\pklab\geq 0$ and $\qklab\geq 0$ for any combination of $k,l,a,b$.
When $\tklab=1$,  $\Pskone$ and $\Psltwo$ take the value of $\pklab$ and $\qklab$, respectively.
\end{itemize}

Let us collect all indicator and power variables in the sets $\Iset$
and $\Pset$, respectively,
and define $\RA = \{\Iset,\Pset\}$.
Every feasible RA scheme can be described by
an $\RA$ satisfying simultaneously
\begin{align}
&\tklu,\tklab\in \{0,1\}, \forall\;k,l,u,a,b,                              \label{eq:first-constr}\\
&\sum_{l}\left(\sum_{u}\tklu + \sum_{a,b}\tklab\right) = 1, \forall\;k,   \label{eq:second-constr}\\
&\sum_{k}\left(\sum_{u}\tklu + \sum_{a,b}\tklab\right) = 1, \forall\;l,   \label{eq:third-constr}\\
&\sum_{k,l,u,a,b} \left(\tklu\Pklu + \tklab(\pklab+\qklab)\right) \leq \Ptot, \label{eq:forth-constr}\\
&\Pklu\geq0, \pklab\geq0,\qklab\geq0,\forall\;k,l,u,a,b,        \label{eq:fifth-constr}\\
&\Pklu=0 {\rm\;if\;}\tklu=0, \forall\;k,l,u,a,b,                              \label{eq:six-sixconstr}\\
&\pklab=0, \qklab=0 {\rm\;if\;}\tklab=0, \forall\;k,l,u,a,b,         \label{eq:last-constr}
\end{align}
where \eqref{eq:second-constr} and \eqref{eq:third-constr} guarantee the OFDMA,
i.e., every subcarrier is used exclusively for the transmission of message bits to a unique user.
\eqref{eq:forth-constr} and \eqref{eq:fifth-constr} ensure the total power constraint is satisfied.
The constraints \eqref{eq:six-sixconstr} and \eqref{eq:last-constr} are added
to guarantee that every $\RA$ is one-to-one mapped to a new variable
for the change of variable (COV) proposed later to solve the RA problem.

Note that an $\RA$ satisfying \eqref{eq:first-constr}-\eqref{eq:last-constr}
indicates a unique feasible RA scheme for the adopted protocol.
Viewed from the other way around, any feasible RA scheme can also be
described by an $\RA$ satisfying those constraints.
Interestingly, the same feasible RA scheme might be described by
multiple different $\RA$ all satisfying these constraints.
For instance, consider the scenario where there is only a single user $u$,
and the RA scheme requiring messages to be transmitted in the direct mode,
respectively, through subcarriers $k_1$ and $k_2$ during the first slot
and subcarriers $l_1$ and $l_2$ during the second slot.
This RA scheme can be described by using either an $\RA$ with
$t_{k_1l_1uu}=t_{k_2l_2uu}=1$ and $t_{k_1l_2uu}=t_{k_2l_1uu}=0$,
or another $\RA'$ with $t_{k_1l_2uu}=t_{k_2l_1uu}=1$ and $t_{k_1l_1uu}=t_{k_2l_2uu}=0$.

Given a feasible $\RA$, the maximum WSR for the adopted protocol is
\begin{align}
f(\RA) = &\sum_{k,l,u,a,b}\big(\tklu \wu\Rate{\Gklu \Pklu} + \\
         &\hspace{0.2cm}   \tklab\big(\wa\Rate{\Gsak \pklab} + \wb\Rate{\Gsbl\qklab}\big),   \nonumber
\end{align}
where $\wu>0$ is the weight prescribed for user $u$.
The WSR maximization problem is to solve
\begin{align}
{\rm (P1)\hspace{0.8cm}}
\max_{\RA}  \hspace{0.25cm} f(\RA)    \hspace{0.25cm}
{\rm s.t.}  \hspace{0.25cm} \eqref{eq:first-constr}-\eqref{eq:last-constr} \nonumber %\label{prob:firstRA}\\
\end{align}
for a globally optimum $\RA$.
We will develop an algorithm in the following subsections to find it,
after which the optimum subcarrier assignment and source/relay power allocation
can be computed according to the analysis in Section \ref{sec:pair-ratemax}.

\subsection{The idea behind the RA algorithm design}\label{sec:overview}

Note that (P1) is a nonconvex program consisting of
both continuous and binary variables,
thus in general its duality gap is not zero.
Similar nonconvex optimization problems for multicarrier systems
exist in the literature \cite{Yu06,Cioffi06}.
A possible approach to tackle them is to show their duality gaps approach zero
when a sufficiently large number of subcarriers is used.
This justifies the use of the dual method to find an asymptotically optimum solution.

Here, we use a continuous-relaxation based approach
to find at least an approximately optimum $\RA$ for (P1).
Similar methods were also used in \cite{Yu02,Cioffi04}
to compute asymptotic capacity regions.
Specifically, all indicator variables are first relaxed to be continuous
within $[0,1]$, after which we get a new problem
\begin{align}
{\rm (P2)\hspace{0.8cm}}
\max_{\RA}  &\hspace{0.25cm}  f(\RA)    \hspace{0.25cm}                     \nonumber\\
{\rm s.t.}  &\hspace{0.25cm}  \tklu,\tklab\in [0,1], \forall\;k,l,u,a,b,   \label{eq:new-firstconstr}\\
            &\hspace{0.25cm}  \eqref{eq:second-constr}-\eqref{eq:last-constr}, \nonumber %\label{prob:firstRA}\\
\end{align}
as a relaxation of (P1). Define the feasible set of (P2) as $\feasRA$.
Obviously, the feasible set of (P1) is a subset of $\feasRA$.

Then, we make the COV from $\Pset$ to $\newPset=\{\newPklu,\newpklab,\newqklab|\forall k,l,u,a,b\}$,
where every $\newPklu$, $\newpklab$ and $\newqklab$ satisfy, respectively,
\begin{align}
\newPklu = \tklu\Pklu, \newpklab=\tklab\pklab, \newqklab=\tklab\qklab.   \label{eq:mapping}
\end{align}

After the COV, we collect all variables into $\newRA = \{\Iset,\newPset\}$.
It is important to note that an $\RA\in\feasRA$ is one-to-one mapped
to an $\newRA\in\feasnewRA$, where $\feasnewRA$ contains the set of
all $\newRA$'s satisfying \eqref{eq:new-firstconstr},
\eqref{eq:second-constr}-\eqref{eq:third-constr}, as well as
\begin{align}
&\sum_{k,l,u,a,b} \left(\newPklu + \newpklab + \newqklab\right) \leq \Ptot, \label{eq:new-forthconstr}\\
&\newPklu\geq0, \newpklab\geq0,\newqklab\geq0,\forall\;k,l,u,a,b,    \label{eq:new-fifthconstr}\\
&\newPklu=0 {\rm\;if\;}\tklu=0, \forall\;k,l,u,a,b,             \label{eq:new-sixconstr}\\
&\newpklab=0, \newqklab=0 {\rm\;if\;}\tklab=0, \forall\;k,l,u,a,b.          \label{eq:new-lastconstr}
\end{align}

As a function of $\newRA\in\feasnewRA$, the WSR can be rewritten as
\begin{align}
g(\newRA) =&  f(\RA(\newRA))       \nonumber \\
          =& \sum_{k,l,u,a,b} \big(\wu\phi(\tklu,\newPklu,\Gklu)  \label{eq:new-WSR}\\
           &  + \wa\phi(\tklab,\newpklab,\Gsak) + \wb\phi(\tklab,\newqklab,\Gsbl)\big), \nonumber
\end{align}
where $\RA(\newRA)$ represents the $\RA$ corresponding to
the $\newRA\in\feasnewRA$ and
\begin{align}
\phi(t,x,G) = \left\{\begin{array}{ll}
                     t \, \Rate{G \, \frac{x}{t}}   &   {\rm if\;}t>0,  \\
                     0                              &   {\rm if\;}t=0.  \\
                   \end{array}\right.
\end{align}

It can readily be shown that $\phi(t,x,G)$ with fixed $G$
is a continuous and concave function of $t\geq 0$ and $x$,
because it is a perspective function of $\Rate{G x}$
which is concave of $x$ (see pages $89-90$ for more details in \cite{Convex-opt}).
Therefore, $g(\newRA)$ is a concave function of $\newRA\in\feasnewRA$.

After solving
\begin{align}
{\rm (P3)\hspace{0.8cm}}
\max_{\newRA} & \hspace{0.25cm} g(\newRA)     \nonumber\\
{\rm s.t.}    & \hspace{0.25cm} \eqref{eq:new-firstconstr},\eqref{eq:second-constr}-\eqref{eq:third-constr},
                \eqref{eq:new-forthconstr}-\eqref{eq:new-lastconstr}, \nonumber
\end{align}
for its global optimum, the $\RA$ corresponding to this global optimum
is the optimum solution for (P2).
In the following subsection, we will focus on solving the problem
\begin{align}
{\rm (P4)\hspace{0.8cm}}
\max_{\newRA} & \hspace{0.25cm} g(\newRA)    \nonumber\\
{\rm s.t.}    & \hspace{0.25cm} \eqref{eq:new-firstconstr},\eqref{eq:second-constr}-\eqref{eq:third-constr},
                \eqref{eq:new-forthconstr}-\eqref{eq:new-fifthconstr}, \nonumber
\end{align}
which is a relaxation of (P3) by omitting \eqref{eq:new-sixconstr}
and \eqref{eq:new-lastconstr}.
Obviously, $\feasnewRA$ is a subset of the feasible set of (P4).
Most interestingly, (P4) is a convex program,
which can be solved by highly-efficient convex-optimization techniques.
Define the optimum objective value for (P1) and (P4) as
$f^\star$ and $g^\star$, respectively.
According to the relaxations we made,
\begin{align}
g^\star \geq \max_{\newRA\in\feasnewRA}g(\newRA)
        = \max_{\RA\in\feasRA}f(\RA) \geq f^\star       \nonumber
\end{align}
follows. Define a global optimum for (P4) as $\newRA^\star$.
If we can find an $\newRA^\star$ that satisfies \eqref{eq:new-sixconstr}
and \eqref{eq:new-lastconstr}, and contains binary indicator variables
(i.e., $\tklu,\tklab\in\{0,1\},\forall\;k,l,u,a,b$),
then it can readily be shown that $\RA(\newRA^\star)$
must be a global optimum for (P1).

In practice, it may be difficult to find precisely
a global optimum $\newRA^\star$ for (P4) in general.
For instance, existing convex-optimization techniques such as
the interior-point method or the dual method all search for the global optimum
in an iterative manner, and finally produce an approximately optimum solution
with an objective value very close to the optimum value.
Motivated by this fact, suppose a solution $\newRA'$ which satisfies
\begin{enumerate}
\item
\eqref{eq:new-sixconstr} and \eqref{eq:new-lastconstr}
and all indicator variables in $\newRA'$ are binary;
\item
$g^\star - g(\newRA')$ is very small;
\end{enumerate}
can be found for (P4),
then $\RA(\newRA')$ is feasible for (P1) and
$f^\star - f(\RA(\newRA'))$ is also very small because
\begin{align}
f^\star - f(\RA(\newRA')) \leq g^\star - g(\newRA'), \nonumber
\end{align}
which means that $\RA(\newRA')$ can be taken as an approximately optimum solution for (P1).

In the following subsection, we use the dual method to solve (P4).
Specifically, the ellipsoid method is used to search for the dual optimum.
This ellipsoid method is reduced to the bisection method to update upper
and lower bounds for the dual optimum iteratively until convergence.
In some cases, the global optimum for (P1) can be found,
while in other cases we explain by theoretical analysis and
illustrate by numerical experiments that,
the optimum solution for the Lagrangian relaxation problem (LRP) of (P4)
corresponding to the upper bound produced after convergence
can be taken as the $\newRA'$ described above.
Then, $\RA(\newRA')$ can be output as an approximately optimum solution for (P1).

\subsection{The development of the RA algorithm}

Since (P4) is a convex program and it satisfies the Slater constraint
qualification\footnote{There exists at least an $\newRA$ satisfying all
inequality constraints strictly.},
(P4) has zero duality gap (see page 226 of \cite{Convex-opt}),
which justifies the use of the dual method to solve (P4).
To this end, $\mu$ is introduced as a Lagrange multiplier
for the constraint \eqref{eq:new-forthconstr}.
The LRP for (P4) is
\begin{align}
{\rm(P5)\hspace{1cm}}
\max_{\newRA}  &\hspace{0.25cm} L(\mu,\newRA) = g(\newRA) + \mu\bigg(\Ptot - \sumP(\newRA)\bigg)   \nonumber\\
{\rm s.t.}     &\hspace{0.25cm} \eqref{eq:new-firstconstr},\eqref{eq:second-constr}-\eqref{eq:third-constr},
                       \eqref{eq:new-fifthconstr},  \nonumber
\end{align}
where $L(\mu,\newRA)$ is the Lagrangian of (P4) and
$\sumP(\newRA)$ is the left-hand side of \eqref{eq:new-forthconstr}
(i.e., the sum power as a function of $\newRA$).
A global optimum for (P5) is denoted by $\newRA_\mu$.
The dual function is defined as $d(\mu)=L(\mu,\newRA_\mu)$,
which is a convex function of $\mu$. In particular,
\begin{align}
\subgrad(\mu) = \Ptot-\sumP(\newRA_\mu)
\end{align}
is a subgradient of $d(\mu)$, i.e., it satisfies
\begin{align}
\forall\;\mu', d(\mu') \geq d(\mu) + (\mu'-\mu)\subgrad(\mu),   \label{eq:property-subgrad}
\end{align}
and the dual problem is to find the dual optimum
\begin{align}
\mu^\star = \arg\min_{\mu\geq0}d(\mu).   \label{prob:dual}
\end{align}

Since (P4) has zero duality gap, the following properties hold:
\begin{itemize}
\item
Note that $\mu^\star$ represents the sensitivity\footnote{
Note that the sensitivity analysis
was introduced in pages 249-253 of \cite{Convex-opt} for a convex minimization problem.
It can be proven that $\frac{g(\newRA^\star)}{\Ptot} = \mu^\star$
by casting the problem (P4) into an equivalent convex minimization problem.
The proof is straightforward and omitted here due to space limitation.}
of the optimum objective value for (P4) with respect to $\Ptot$,
i.e., $\frac{g(\newRA^\star)}{\Ptot} = \mu^\star$.
Obviously, $g(\newRA^\star)$ is strictly increasing of $\Ptot$, meaning that $\mu^\star>0$.

\item
$\mu=\mu^\star$ and $\newRA_{\mu} = \newRA^\star$ are true
if and only if $\newRA_\mu$ is feasible and $\mu \subgrad(\mu) = 0$ is satisfied
according to Proposition $5.1.5$ in \cite{Nonlinear-opt}.
This means that $\mu^\star\subgrad(\mu^\star)=0$.
Moreover, $\newRA_{\mu} = \newRA^\star$ if $\subgrad(\mu)=0$.
\end{itemize}

The idea behind the dual method to solve (P4) is to search for $\mu^\star$.
Then, the $\newRA_{\mu^\star}$ that satisfies $\subgrad(\mu^\star) = 0$
can be taken as $\newRA^\star$.
The key to the dual method consists of two procedures
to find $\newRA_{\mu}$ for a given $\mu>0$ and $\mu^\star$, respectively,
which are developed as follows.

\subsubsection{Finding $\newRA_\mu$ when $\mu>0$}\label{sec:solve-LRP}

The following strategy is used to find $\newRA_\mu$ for (P5) when $\mu>0$.
First, the optimum $\newPset$ for (P5) with fixed $\Iset$ is found and
denoted by $\newPset_\Iset$.
Define $\newRA_\Iset = \{\Iset, \newPset_\Iset\}$.
Then we find the optimum $\Iset$ to maximize $L(\mu,\newRA_\Iset)$
subject to \eqref{eq:new-firstconstr}, \eqref{eq:second-constr} and \eqref{eq:third-constr}.
Finally, $\newRA_\Iset$ corresponding to this optimum $\Iset$ can be taken as $\newRA_\mu$.

Suppose $\Iset$ is fixed, we find $\newPset_\Iset$ as follows.
Specifically, every $\newPklu$ in $\newPset_\Iset$ is equal to $0$
when $\tklu=0$. When $\tklu>0$, the optimum $\newPklu$ can be found by
using the KKT conditions related to $\newPklu$.
In summary, the optimum $\newPklu$ can be shown to be
\begin{align}
\newPklu = \tklu \Lambda(\wu,\mu,\Gklu), \label{eq:optPklu}
\end{align}
where $\Lambda(\wu,\mu,G)$ is defined as
$\Lambda(\wu,\mu,G) = \left[\frac{\wu\log_2{e}}{2\mu} - \frac{1}{G}\right]^+$.
In a similar way, the optimum $\newpklab$ and $\newqklab$
can be shown to be
\begin{align}
\newpklab &= \tklab\Lambda(\wa,\mu,G_{sa,k}),  \label{eq:optpklab}\\
\newqklab &= \tklab\Lambda(\wb,\mu,G_{sb,l}),  \label{eq:optqklab}
\end{align}
respectively.
Using these formulas, $\newRA_\Iset = \{\Iset, \newPset_\Iset\}$
can be found. It can readily be shown that
\begin{align}
L(\mu,\newRA_\Iset) = \mu\Ptot + \sum_{k,l,u,a,b} \big(\tklu\Aklu + \tklab\Bklab\big)
\end{align}
where
\begin{align}
\Aklu =& \wu\Rate{\Gklu\Lambda(\wu,\mu,\Gklu)} - \mu\cdot\Lambda(\wu,\mu,\Gklu)   \nonumber\\
\Bklab = &\wa\Rate{\Gsak\Lambda(\wa,\mu,\Gsak)}- \mu\cdot\Lambda(\wa,\mu,\Gsak) +  \nonumber \\
         &\wb\Rate{\Gsbl\Lambda(\wb,\mu,\Gsbl)}- \mu\cdot\Lambda(\wb,\mu,\Gsbl).    \nonumber
\end{align}

Finally, we find the optimum $\Iset$ for maximizing $L(\mu,\newRA_\Iset)$ subject to
\eqref{eq:new-firstconstr}, \eqref{eq:second-constr} and \eqref{eq:third-constr}.
This problem is equivalent to solving
\begin{align}
\max_{\Iset,\{\tkl|\forall\;k,l\}}
&\hspace{0.25cm} \sum_{k,l}\sum_{u,a,b} \big(\tklu\Aklu + \tklab\Bklab\big)   \nonumber\\
{\rm s.t.}    &\hspace{0.25cm} \sum_{l}\tkl = 1, \forall\;k,                     \label{prob:find-Ikl-first}\\
              &\hspace{0.25cm} \sum_{k}\tkl = 1, \forall\;l,                     \nonumber\\
              &\hspace{0.25cm} \tkl = \sum_{u}\tklu + \sum_{a,b}\tklab,\forall\;k,l.       \nonumber\\
              &\hspace{0.25cm} \tklu\geq 0, \tklab\geq0,\forall\;k,l,u,a,b. \nonumber
\end{align}

Note that the inequality $\sum_{u,a,b} \big(\tklu\Aklu + \tklab\Bklab\big) \leq \tkl \Ckl$
holds where $\Ckl = \max\{\max_{u}\Aklu, \max_{a,b}\Bklab\}$.
Let us call $\Aklu$ as the metric for $\tklu$ and $\Bklab$ as the metric for $\tklab$.
This inequality is tightened when all entries of $\{\tklu,\tklab|\forall\;u,a,b\}$
are assigned to zero, except that the one with the metric equal to $\Ckl$ is assigned to $\tkl$.

Therefore, after the problem
\begin{align}
\max_{\{\tkl|\forall\;k,l\}}  &\hspace{0.25cm} \sum_{k,l}\sum_{u,a,b} \tkl\Ckl  \nonumber\\
{\rm s.t.}    &\hspace{0.25cm} \sum_{l}\tkl = 1, \forall\;k,                     \label{prob:find-Ikl-second}\\
              &\hspace{0.25cm} \sum_{k}\tkl = 1, \forall\;l,                     \nonumber\\
              &\hspace{0.25cm} \tkl \geq 0, \forall\;k,l,   \nonumber
\end{align}
is solved for its optimum solution $\{\tkl^\star|\forall\;k,l\}$,
an optimum $\Iset$ for \eqref{prob:find-Ikl-first} can be constructed by assigning
for every combination of $k$ and $l$, all entries in $\{\tklu,\tklab|\forall\;u,a,b\}\subset\Iset$
to zero, except for the one with the metric equal to $\Ckl$ to $\tkl^\star$.

Most interestingly, \eqref{prob:find-Ikl-second} is a standard assignment problem,
hence every entry in $\{\tkl^\star|\forall\;k,l\}$ is either $0$ or $1$
and $\{\tkl^\star|\forall\;k,l\}$ can be found efficiently by the Hungarian algorithm \cite{Hungarian}.
After knowing $\{\tkl^\star|\forall\;k,l\}$, the optimum $\Iset$
can be constructed according to the way mentioned earlier.
Finally, the corresponding $\newRA_\Iset = \{\Iset,\newPset_\Iset\}$ is assigned to $\newRA_\mu$.
Note that to compute $\newRA_\mu$, $\{\Aklu,\Bklab|\forall\;k,l,u,a,b\}$
containing $K^2(U+U^2)$ entries has to be computed first, which implies a complexity of $O(K^2U^2)$.
Moreover, the Hungarian algorithm to solve \eqref{prob:find-Ikl-second}
has a complexity of $O(K^3)$ \cite{Hungarian}.
This means that the complexity of finding $\newRA_\mu$ is $O(K^2U^2 + K^3)$.

\subsubsection{Finding $\mu^\star$}

To find $\mu^\star$, an incremental-update based subgradient method
which updates $\mu$ with $\mu = [\mu - \delta(\Ptot-\sumP(\newRA_\mu))]^+$ can be used,
where $\delta>0$ is a prescribed step size \cite{Nonlinear-opt}.
However, this method converges very slowly,
since $\delta$ has to be very small to guarantee convergence.
To speed up the search for $\mu^\star$, we use the ellipsoid method.
The idea behind the ellipsoid method is to find a series of
contracting ellipsoids that always contain $\mu^\star$ \cite{Convex-opt}.
The ellipsoid method can be reduced to the bisection method as follows.

First, a lower bound $\mumin$ and an upper bound $\mumax$
for $\mu^\star$ are initialized.
As said earlier, $\mu^\star>0$ holds, thus $\mumin$ can be initialized with $0$.
As shown in the Appendix,
$\mumax$ can be initialized with $\frac{K\wmax\log_2{e}}{\Ptot}$.
Then, $\mumin$ and $\mumax$ are updated iteratively as follows.
In every iteration, $\newRA_{\mumid}$ where $\mumid=\frac{\mumin+\mumax}{2}$
is computed.
If $\subgrad(\mumid) > 0$, then $\forall\;\mu > \mumid$,
$d(\mu) \geq d(\mumid) + (\mu-\mumid)\subgrad(\mumid) > d(\mumid)$.
This means that $\mu^\star$ must be confined in $[\mumin, \mumid]$,
so $\mumax$ should be updated with $\mumid$.
If $\subgrad(\mumid) < 0$, it can be shown similarly that
$\mumin$ should be updated with $\mumid$.
The iteration is terminated when $\subgrad(\mumid) = 0$
or $\mumax-\mumin\leq \epsilon$ where $\epsilon>0$ is a prescribed small value.

When the iteration is terminated with $\subgrad(\mumid) = 0$ being satisfied,
$\newRA^\star=\newRA_{\mumid}$ must hold as said earlier.
Note that $\RA(\newRA_{\mumid})$ must be a global optimum for (P1)
since $\newRA_{\mumid}$ satisfies \eqref{eq:new-sixconstr} and \eqref{eq:new-lastconstr},
and contains binary indicator variables as said in Section IV.B.

We now consider the case where the iteration is terminated
with $\mumax-\mumin\leq \epsilon$ being satisfied.
In such a case, we find that $\newRA_{\mumax}$ is
an approximately optimum solution for (P4).
This finding will be illustrated by numerical experiments in Section V.
It can be explained by theoretical analysis as follows.
Note that
\begin{align}
g^\star - g(\newRA_{\mumax}) \leq d(\mumax) - g(\newRA_{\mumax})
= \mumax \subgrad(\mumax)  \label{eq:ineqal}
\end{align}
holds since $\forall\;\mu\geq0$, $g^\star\leq d(\mu)$.
In addition, we present the following lemma:

\begin{lemma}
$\subgrad(\mu)$ is an increasing function of $\mu\geq0$.
\end{lemma}
\begin{IEEEproof}
Suppose $\mu_1\geq \mu_2$.
According to \eqref{eq:property-subgrad},
\begin{align}
d(\mu_1) &\geq  d(\mu_2) + (\mu_1-\mu_2)\subgrad(\mu_2)   \nonumber\\
d(\mu_2) &\geq  d(\mu_1) + (\mu_2-\mu_1)\subgrad(\mu_1)  \nonumber
\end{align}
follow. As a result,
\begin{align}
(\mu_1 - \mu_2)\subgrad(\mu_1) \geq  d(\mu_1) - d(\mu_2)
   \geq  (\mu_1-\mu_2)\subgrad(\mu_2)  \nonumber
\end{align}
holds, and thus $\subgrad(\mu_1) \geq \subgrad(\mu_2)$. This completes the proof.
\end{IEEEproof}

According to Lemma 1, $\subgrad(\mumax)\geq \subgrad(\mu^\star)=0$
because $\mumax\geq \mu^\star$,
meaning that $\newRA_\mumax$ is always feasible for (P4).
Moreover, $\mumax\subgrad(\mumax)$ reduces as the iteration proceeds
and it is very small after convergence,
since $\mumax$ decreases to approach $\mu^\star$
which satisfies $\mu^\star\subgrad(\mu^\star)=0$.
This means that $g^\star - g(\newRA_{\mumax})$
is very small according to \eqref{eq:ineqal}.
Moreover, $\newRA_{\mumax}$ also satisfies
\eqref{eq:new-sixconstr} and \eqref{eq:new-lastconstr}
and all indicator variables in $\newRA_{\mumax}$ are binary.
This means that $\RA(\newRA_{\mumax})$ can be output as
an approximately optimum solution for (P1) as said in Section IV.B.

The overall procedure to find an approximately optimum solution for (P1)
is summarized in Algorithm \ref{alg:dual}.
Its complexity can be studied as follows.
First, $\{\Gklu|\forall\;k,l,u\}$ needs to be computed, which needs $K^2U$ operations.
Then, finding $\mu^\star$ with the bisection method
requires at most a number of iterations in the order of $\log_2(K)$.
For each iteration, computing $\newRA_\mu$ has a complexity of $O(K^2U^2 + K^3)$.
Therefore, the total complexity of
Algorithm \ref{alg:dual} is $O(\log_2(K)(K^2U^2+K^3))$.

\begin{algorithm}
\caption{The RA algorithm to find an approximately optimum $\RA$ for (P1)} \label{alg:dual}
\begin{algorithmic}[1]
\STATE  compute $\Gklu$, $\forall\;k,l,u$.
\STATE  $\mumin = 0$; $\mumax = \frac{K\wmax\log_2{e}}{\Ptot}$;
\WHILE{$\mumax - \mumin > \epsilon$}
       \STATE  $\mumid = \frac{\mumax+\mumin}{2}$;
       \STATE  solve (P5) with $\mu=\mumid$ for $\newRA_{\mumid}$;
               compute $\subgrad(\mumid)$;
       \IF{$\subgrad(\mumid) = 0$}
           \STATE  compute $\RA(\newRA_{\mumid})$
                   and output it as an optimum solution for (P1);
           \STATE  exit the algorithm;
       \ELSIF{$\subgrad(\mumid) > 0$}
           \STATE  $\mumax = \mumid$;
       \ELSE
           \STATE  $\mumin = \mumid$;
       \ENDIF
\ENDWHILE

\STATE  solve (P5) with $\mu=\mumax$ for $\newRA_{\mumax}$;
\STATE  compute $\RA(\newRA_{\mumax})$ and output it
        as an approximately optimum solution for (P1).
\end{algorithmic}
\end{algorithm}

\section{Numerical experiments}\label{sec:numexp}

\begin{figure}[h]
  \centering
     \includegraphics[width=3.5in]{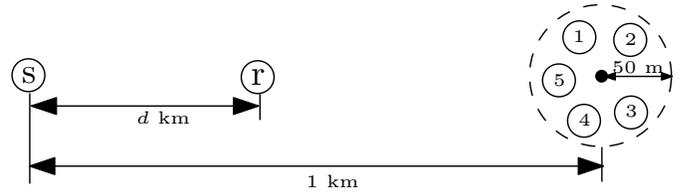}
  \caption{The relay-aided downlink OFDMA system considered in numerical experiments.}  \label{fig:simsys}
\end{figure}

In numerical experiments, we consider the relay-aided downlink OFDMA system
illustrated in Figure \ref{fig:simsys}.
The relay is located in the line between the source and the center
of the user region, and the source-to-relay distance is $d$ km.
$U=5$ users are served and they are randomly and uniformly distributed
in a circular region of radius $50$ m.
Their weights are randomly chosen between $0.8$ and $1.2$
for every system realization simulated.
For Algorithm 1, $\epsilon$ is set as $10^{-6}$,
which leads to at most
$\log_2(\frac{K\wmax\log_2{e}}{\epsilon\Ptot})\approx 21+\log_2(\frac{K}{\Ptot})$
iterations for a given combination of $K$ and $\Ptot$.

The channels are independent of each other and generated in the same
way as in \cite{Wang11TSP-1,Wang12TSP}.
For every user $u$, the impulse response of the source-to-$u$ channel
is modeled as a delay line with $L=6$ taps, which are independently generated from
circularly symmetric complex Gaussian distributions with zero mean and variance
equal to $\frac{1}{L}\big(\frac{\dsu}{\dref}\big)^{-2.5}$,
where $\dref=1$ km and $\dsu$ represents the source-to-$u$ distance.
The source-to-relay and relay-to-$u$ channels are generated in the same way,
with each tap having the variance as
$\frac{1}{L}\big(\frac{d}{\dref}\big)^{-2.5}$ and
$\frac{1}{L}\big(\frac{\dru}{\dref}\big)^{-2.5}$, respectively,
where $\dru$ represents the relay-to-$u$ distance.
The CSI $\{\hsrk|\forall\;k\}$, $\{\hsuk|\forall\;k,u\}$ and $\{\hruk| \forall\;k,u\}$
are computed by making $K$-point FFT over the impulse response of the associated channels.

In order to illustrate the benefit of optimized subcarrier pairing
and opportunistic DF relaying, we also consider
another benchmark protocol (BP-2) in addition to
the already studied benchmark mark protocol (BP-1).
BP-2 is the one studied in \cite{WangTSP11} using a single relay,
i.e., subcarrier $k$ in the first slot and subcarrier $k$ in the second slot
are allocated to a user for either the relay-aided transmission
or the direct transmission separately.
The RA algorithm proposed in \cite{WangTSP11} is used for BP-2.

According to the analysis in Section IV.C,
$\RA(\newRA_{\mumax})$ is finally output as an approximately optimum solution
if the iteration is terminated with $\mumax-\mumin\leq \epsilon$ being satisfied.
In such a case, $f^\star - f(\RA(\newRA_{\mumax})) \leq \mumax\subgrad(\mumax)$
after convergence, and
\begin{align}
\delta(\mumax) = \frac{\mumax\subgrad(\mumax)}{f(\RA(\newRA_{\mumax}))}
\end{align}
can be computed to evaluate the relative difference between
the WSR finally achieved and the maximum WSR for (P1).

\begin{figure}[h]
  \centering
     \includegraphics[width=3.5in,height=2in]{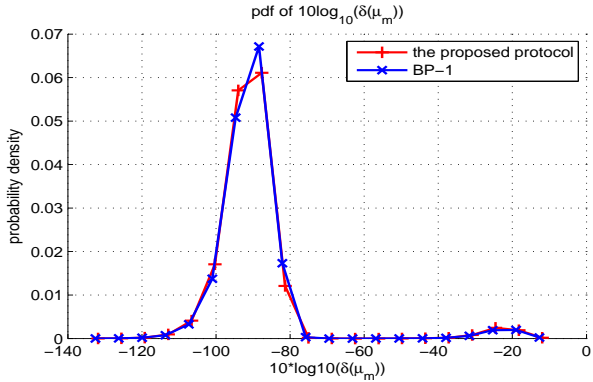}
  \caption{The PDF of $10*log10(\delta(\mumax)$ simulated over $10^4$
           random system realizations.}  \label{fig:err-pdf}
\end{figure}

To illustrate the effectiveness of Algorithm 1,
we have executed Algorithm 1 for both the proposed protocol and BP-1
over $10^4$ random system realizations.
Specifically, the system realizations are generated by
randomly choosing a combination of $d\in[0.1,0.9]$ km, $K\in\{8, 16, 32, 64, 128\}$,
$\Ptot/\sigma^2\in[0,45]$ dB, then generating the channels as said earlier.
It can readily be shown that at most $28$ iterations are executed
for Algorithm 1 for every random channel realization generated.
The $\delta(\mumax)$ is evaluated and collected
for all system realizations when the iteration of Algorithm 1
terminates with $\mumax-\mumin\leq \epsilon$ being satisfied.
The probability density function (PDF) of these $\delta(\mumax)$
in dB scale (i.e., 10*log$_{10}(\delta(\mumax)$)) is shown in Figure \ref{fig:err-pdf}.
It can be seen that $\delta(\mumax)$ is always smaller than $3\%$,
which indicates that the finally produced
$\RA(\newRA_{\mumax})$ is indeed an approximately optimum solution
with a WSR very close to the maximum WSR for (P1)
if the iteration is terminated with $\mumax-\mumin\leq \epsilon$ being satisfied.

%\subsection{Impact of relay position on the protocols' performance}

\begin{figure}[h]
  \centering
  \subfigure[]{
     \includegraphics[width=3.5in,height=2in]{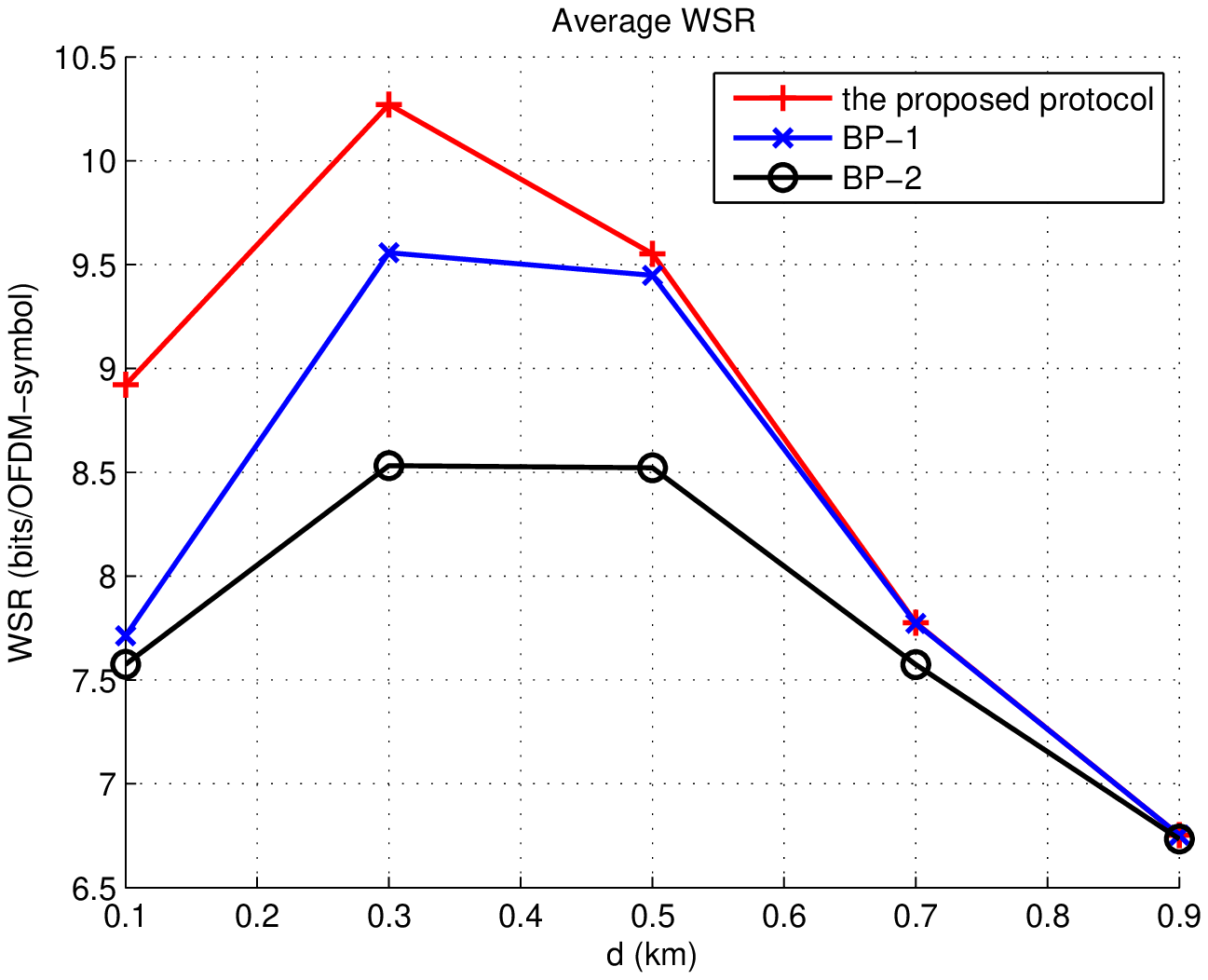}}
  \subfigure[]{
     \includegraphics[width=3.5in,height=2in]{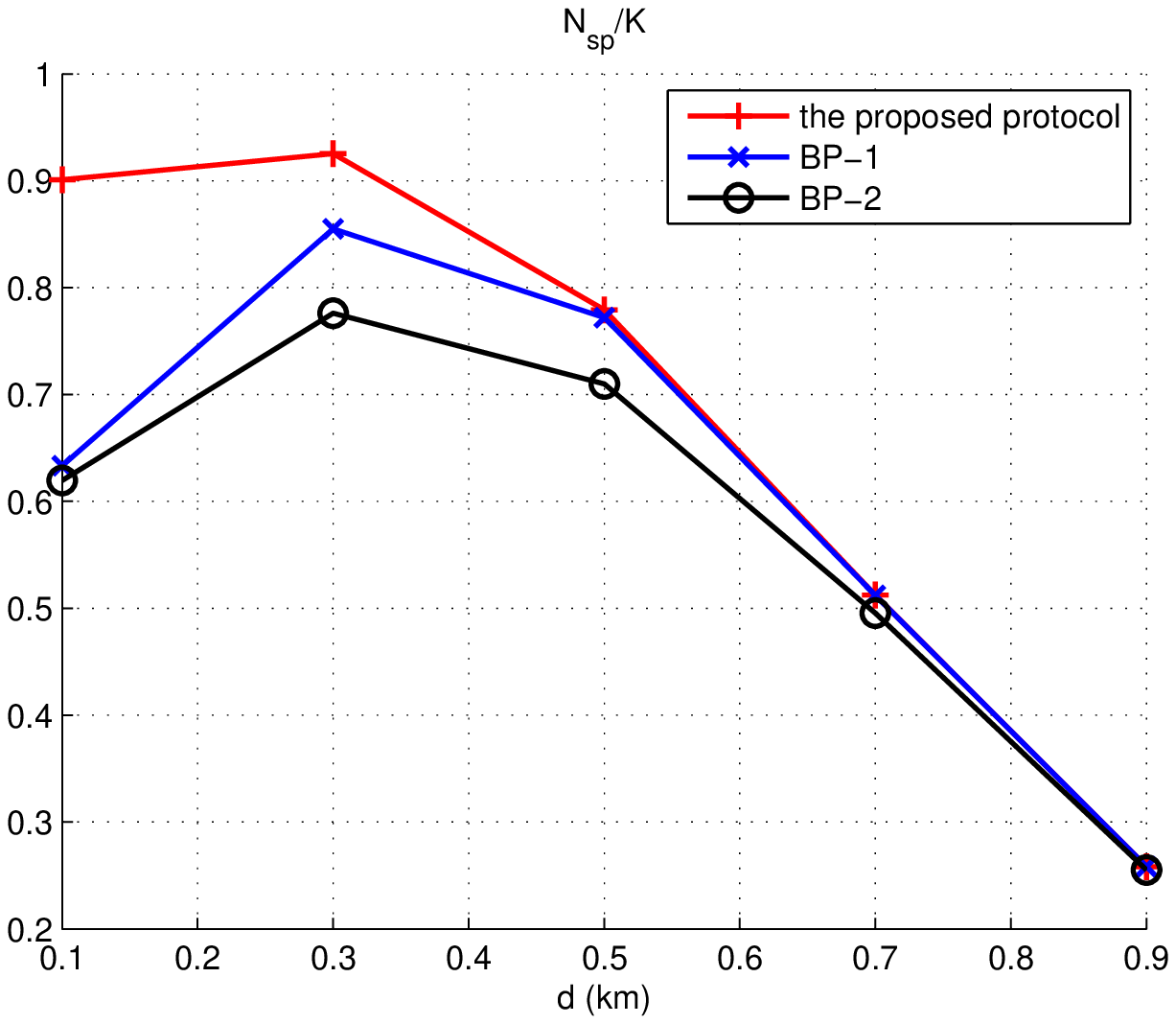}}
  \caption{The average optimum WSRs and $\frac{\Nsp}{K}$
           as the relay position changes when $\Ptot/\sigma^2=20$ dB and $K=32$.}  \label{fig:Pt-20dB}
\end{figure}

To show the impact of relay position on the protocols' performance,
we choose $\Ptot/\sigma^2 = 20$ dB and $K=32$,
then evaluated the average optimum WSRs and $\frac{N_{\rm sp}}{K}$
for every protocol over $1000$ random channel realizations
when $d$ increases from $0.1$ to $0.9$ km.
Here, $\Nsp$ denotes the average number of the subcarrier pairs
that should be used in the relay-aided mode to maximize the WSR.
It can readily be computed that at most $20$ iterations is executed
for Algorithm 1 for every channel realization generated.
The results are shown in Figure \ref{fig:Pt-20dB}.

When $d$ is fixed, the proposed protocol leads to a greater average optimum WSR than BP-1,
which illustrates the theoretical analysis in Section \ref{sec:comparison}.
Moreover, the proposed protocol and BP-1 both have
greater average optimum WSRs than BP-2.
This is because they can better exploit the degrees of freedom
for subcarrier pairing and assignment to users than BP-2
to improve the spectrum efficiency.

It is interesting to observe that for every protocol, the optimum WSR
is higher and it is more likely to pair subcarriers for the relay-aided transmission
to maximize the WSR when the relay moves toward the middle
between the source and the user-region center.
This behavior is interpreted for the proposed protocol as follows
(those for BP-1 and BP-2 can be interpreted in a similar way and thus omitted due to space limitation).
It is important to note that the optimum WSR for the proposed protocol,
as the optimum objective value of (P1),
depends on $\{\Gsuk,\newGklu| \forall\;k,l,u\}$.
If $\forall\;k,l,u$, $\newGklu$ is more likely to take a high value,
the subcarriers are more likely to be paired for the relay-aided transmission
to maximize the WSR, and the average optimum WSR for the proposed protocol increases.
As can be seen from Fig. \ref{fig:novel-kl}, $\newGklu$ is high
if both $\Gsrk$ and $\sumGul$ are much greater than $\Gsuk$.
When the relay lies in the middle between the source and the user-region center,
both $\Gsrk$ and $\sumGul$ are likely to be much greater than $\Gsuk$,
meaning that $\newGklu$ is likely to be high.
Therefore, the optimum WSR is higher and it is more likely
to pair subcarriers for the relay-aided transmission
when the relay lies in the middle between the source and the user-region center.

When $d$ is small, the optimum WSR for the proposed protocol
is much greater than that for BP-1, and it is more likely to pair subcarriers
for the relay-aided transmission to maximize the WSR
for the proposed protocol than for BP-1. This can be explained as follows.
Note that if $\newGklu-\oldGklu$ is very likely to be high $\forall\;k,l,u$,
the proposed protocol is more likely to pair the subcarriers
for the relay-aided transmission than BP-1.
According to the analysis in Section \ref{sec:comparison},
$\newGklu-\oldGklu$ increases when $\Gsrk$ increases or $\Grul$ reduces.
When $d$ is small, $\Gsrk$ and $\Grul$ are very likely to
be high and small, respectively,
meaning that $\newGklu-\oldGklu$ is very likely to take a high value.
This explains the observation.

\begin{figure}[h]
  \centering
  \subfigure[]{
     \includegraphics[width=3.5in,height=2in]{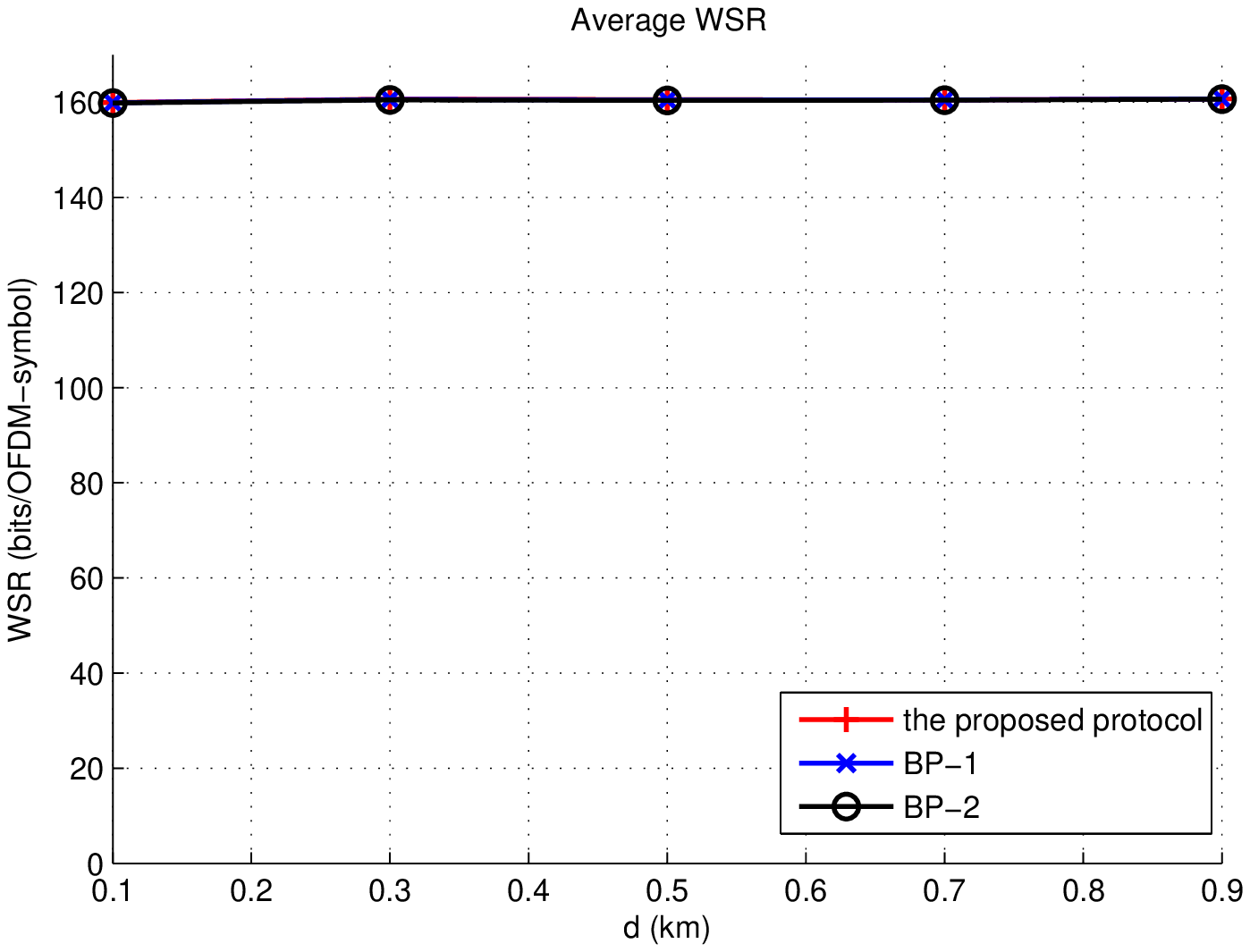}}
  \subfigure[]{
     \includegraphics[width=3.5in,height=2in]{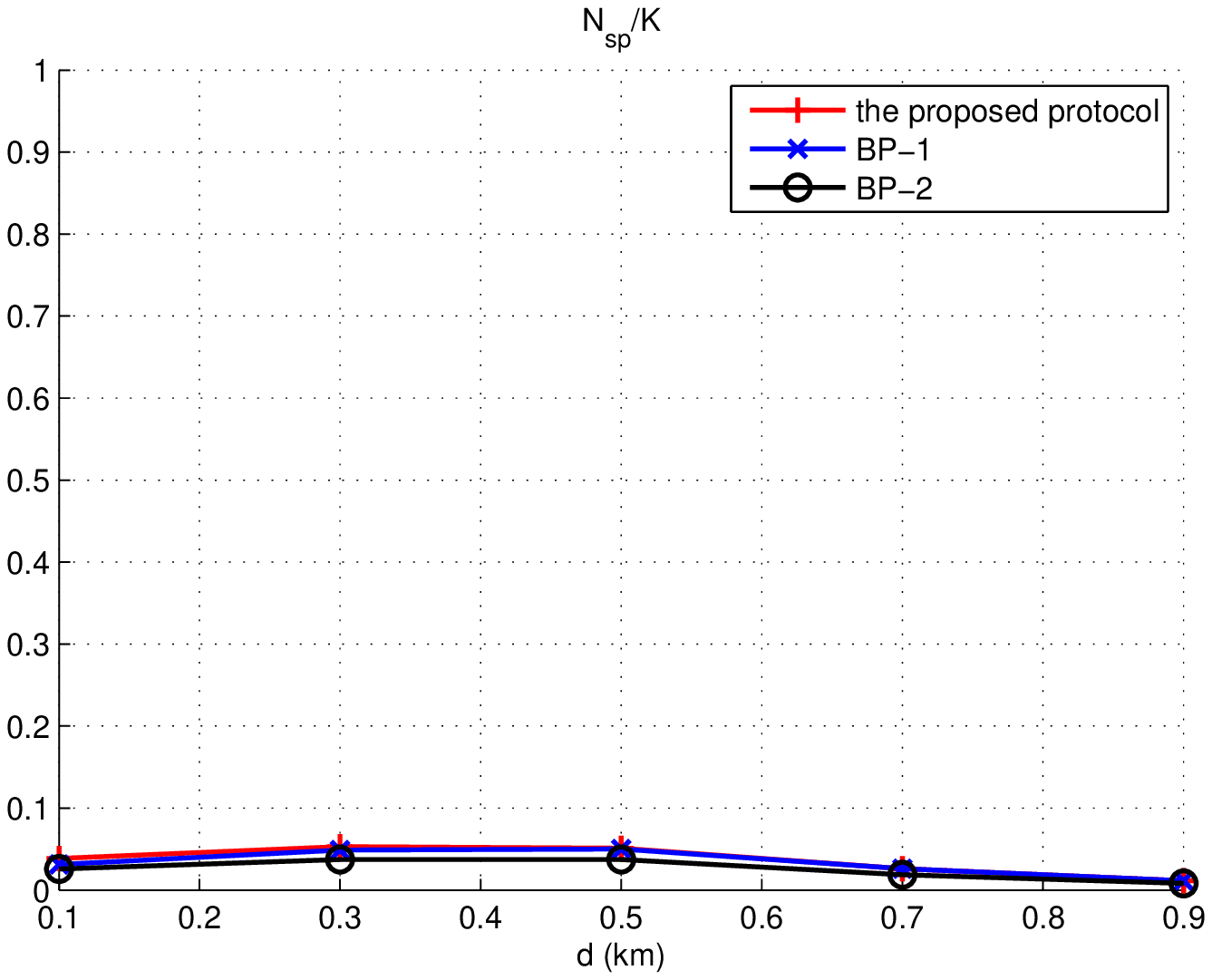}}
  \caption{The average optimum WSRs and $\frac{\Nsp}{K}$
           as the relay position changes when $\Ptot/\sigma^2=45$ dB and $K=32$.}  \label{fig:Pt-40dB}
\end{figure}

We also evaluated the average optimum WSRs and $\frac{N_{\rm sp}}{K}$
for every protocol over $1000$ random channel realizations
when $\Ptot/\sigma^2 = 45$ dB and $K=32$.
It can readily be computed that at most $12$ iterations is executed
for Algorithm 1 for every channel realization generated.
The results are shown in Figure \ref{fig:Pt-40dB}.
It can be seen that regardless of the relay position,
almost all subcarriers are used for the direct transmission to maximize the WSR
for every protocol, therefore all protocols have similar average optimum WSRs.
This can be interpreted as follows.
Note that when the subcarrier pairing and assignment to users are fixed for every protocol,
the optimum sum power for the subcarrier pairs and
the optimum power for unpaired subcarriers can be found by the water-filling method.
Since $\Ptot/\sigma^2$ is very high, the optimum sum power allocated to
subcarriers $k$ and $l$ is very likely to be high
if they are paired for the relay-aided transmission to a user.
In such a case, it can readily be shown that
splitting this high sum power to the two subcarriers
for separate direct transmission to the same user can result in a higher WSR.
This explains why almost all subcarriers are used
for the direct transmission to maximize the WSR when $\Ptot/\sigma^2$ is very high.
It also indicates that the proposed protocol leads to
a better optimum WSR performance than the benchmark ones
especially for the low-power regime.

\section{Conclusion}

In this paper, we have addressed the WSR maximization problem
for the DF relay-aided downlink OFDMA transmission under a total power constraint.
A novel subcarrier-pair based opportunistic DF relaying protocol has been proposed.
A benchmark protocol has also be considered.
An algorithm has been designed to find at least an approximately optimum
RA with a WSR very close to the maximum WSR.
Numerical experiments have illustrated the effectiveness of the RA algorithm
and the impact of relay position and total power on the protocols' performance.
Theoretical analysis have been presented to interpret
what were observed in numerical experiments.

\section*{Acknowledgement}

The authors would like to thank Prof. Wolfgang Utschick
and the anonymous reviewers for their valuable suggestions
to improve the quality of this work.

% if have a single appendix:

\appendix[An upper bound for $\mu^\star$]

An initial upper bound for $\mu^\star$ can be found as follows.
According to Proposition $5.5.1$ in \cite{Nonlinear-opt},
$\newRA^\star$ must satisfy $\newRA_{\mu^\star} = \newRA^\star$ and
$\mu^\star(\Ptot-\sumP(\newRA^\star)) = 0$. 
Since $\mu^\star>0$, $\sumP(\newRA^\star) = \Ptot$ must be satisfied. 
According to the derivation to find $\newRA_{\mu}$, 
the power and indicator variables in $\newRA^\star$ must
satisfy \eqref{eq:optPklu}-\eqref{eq:optqklab}.
It can readily be seen that the $\newPklu^\star$, $\newpklab^\star$
and $\newqklab^\star$ in $\newRA^\star$
are smaller than $\tklu^\star\frac{\wmax\log_2{e}}{2{\mu^\star}}$,
$\tklab^\star\frac{\wmax\log_2{e}}{2{\mu^\star}}$,
and $\tklab^\star\frac{\wmax\log_2{e}}{2{\mu^\star}}$, respectively,
where $\tklu^\star$ and $\tklab^\star$ represent
the value of $\tklu$ and $\tklab$ in $\newRA^\star$.
Therefore, the inequality
\begin{align}
\Ptot = \sumP(\newRA^\star) &\leq \sum_{k,l,u,a,b}
(\tklu^\star + 2\tklab^\star)\frac{\wmax\log_2{e}}{2\mu^\star} \nonumber\\
                  &\leq \sum_{k,l,u,a,b} 2(\tklu^\star + \tklab^\star)\frac{\wmax\log_2{e}}{2\mu^\star} \\
                  &= \frac{K\wmax\log_2{e}}{\mu^\star}  \nonumber
\end{align}
follows where $\wmax = \max_u\{w_u\}$, meaning that
$\mu^\star \leq \frac{K\wmax\log_2{e}}{\Ptot}$ must be satisfied.
Therefore, $\frac{K\wmax\log_2{e}}{\Ptot}$ can be used as an initial upper bound for $\mu^\star$.

%\appendix[Proof of the Zonklar Equations]
% or
%\appendix  % for no appendix heading
% do not use \section anymore after \appendix, only \section*
% is possibly needed

% use appendices with more than one appendix
% then use \section to start each appendix
% you must declare a \section before using any
% \subsection or using \label (\appendices by itself
% starts a section numbered zero.)
%

%\appendices
%\input{Appendix}

%\section{Proof of the First Zonklar Equation}
%Appendix one text goes here.

% Can use something like this to put references on a page
% by themselves when using endfloat and the captionsoff option.
\ifCLASSOPTIONcaptionsoff
  \newpage
\fi

% trigger a \newpage just before the given reference
% number - used to balance the columns on the last page
% adjust value as needed - may need to be readjusted if
% the document is modified later
%\IEEEtriggeratref{8}
% The "triggered" command can be changed if desired:
%\IEEEtriggercmd{\enlargethispage{-5in}}

% references section

% can use a bibliography generated by BibTeX as a .bbl file
% BibTeX documentation can be easily obtained at:
% http://www.ctan.org/tex-archive/biblio/bibtex/contrib/doc/
% The IEEEtran BibTeX style support page is at:
% http://www.michaelshell.org/tex/ieeetran/bibtex/
\bibliographystyle{IEEEtran}
\bibliography{subcarr-pairing}%
% <OR> manually copy in the resultant .bbl file
% set second argument of \begin to the number of references

%\vfill

% Can be used to pull up biographies so that the bottom of the last one
% is flush with the other column.
%\enlargethispage{-5in}

\begin{IEEEbiography}[{\includegraphics[width=1in,height=1.5in,clip,keepaspectratio]{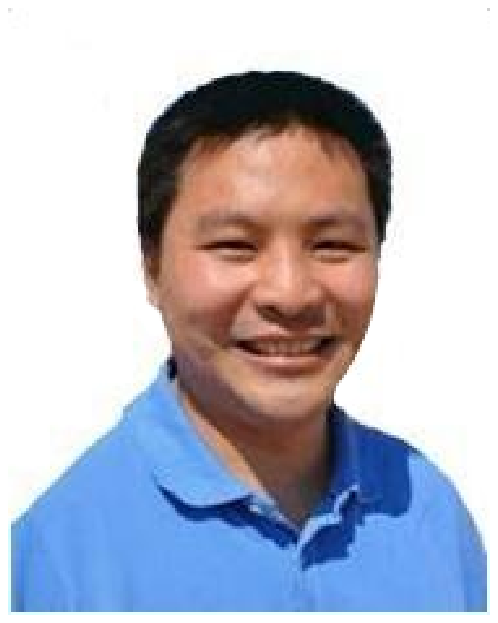}}]{Tao Wang} received respectively Bachelor ({\it summa cum laude}) and Doctor of Engineering degrees
from Zhejiang University, China, in 2001 and 2006, respectively,
as well as Electrical Engineering degree ({\it summa cum laude})
and Ph.D from Universit{\'e} Catholique de Louvain (UCL), Belgium, in 2008 and 2012, respectively.

He has been with School of Communication \& Information Engineering,
Shanghai University, China as a full professor since Feb. 2013,
after winning Professor of Special Appointment (Eastern Scholar) Award
from Shanghai Municipal Education Commission.
Before that, he had multiple research appointments in UCL,
Delft University of Technology and Holst Center in the Netherlands,
Institute for Infocomm Research (I$^2$R) in Singapore,
and Motorola Electronics Ltd. Suzhou Branch in China.
He is an associate editor for {\it EURASIP Journal on Wireless Communications
and Networking} and {\it Signal Processing: An International Journal (SPIJ)},
as well as an editorial board member of {\it Recent Patents on Telecommunications}.
He served as a session chair in 2012 IEEE International Conference on Communications,
and as a TPC member of International Congress on Image and Signal Processing
in 2012 and 2010. He is an IEEE Senior Member.
\end{IEEEbiography}

\begin{IEEEbiography}[{\includegraphics[width=1in,height=1.5in,clip,keepaspectratio]{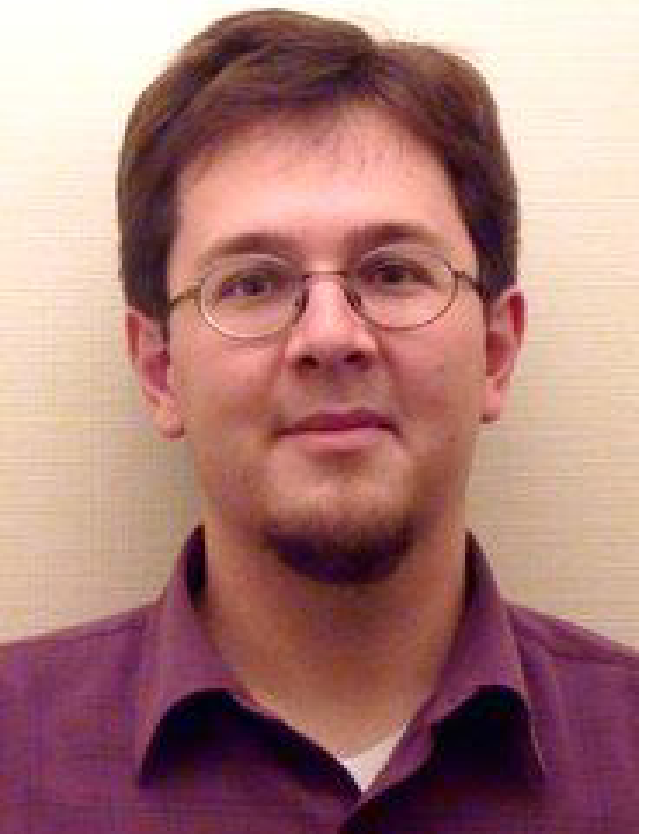}}]
{Fran\c{c}ois Glineur} received engineering degrees from Facult\'e Polytechnique de Mons (Belgium) and \'Ecole sup\'erieure d'\'electricit\'e (France) in 1997,
and a Ph.D. in applied sciences from Facult\'e Polytechnique de Mons in 2001.
He is currently professor of applied mathematics at the Engineering School of
Universit\'e catholique de Louvain, member of the Center for
Operations Research and Econometrics and the Institute of Information
and Communication Technologies, Electronics and Applied Mathematics.
His main research interests lie in convex optimization, including models,
algorithms and applications, as well as in nonnegative matrix factorization.
\end{IEEEbiography}

\begin{IEEEbiography}[{\includegraphics[width=1in,height=1.5in,clip,keepaspectratio]{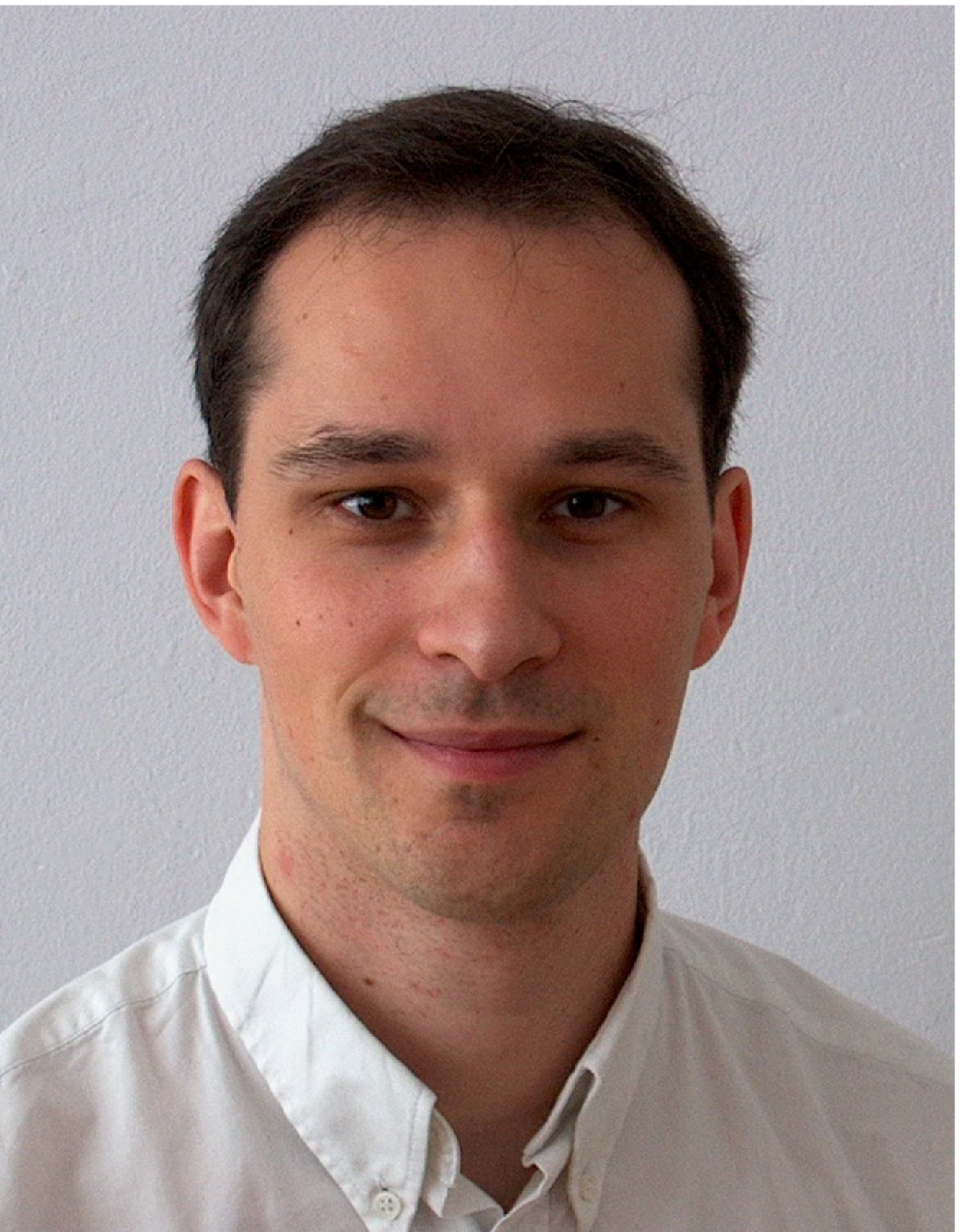}}]
{J\'er{\^o}me Louveaux} received the electrical engineering degree and the
Ph. D. degree from the Universit¨¦ catholique de Louvain (UCL),
Louvain-la-Neuve, Belgium in 1996 and 2000 respectively.

From 2000 to 2001, he was a visiting scholar in the Electrical
Engineering department at Stanford University, CA. From 2004 to 2005, he
was a postdoctoral researcher at the Delft University of technology,
Netherlands. Since 2006, he has been an Associate Professor in the
ICTEAM institute at UCL.  His research interests are in signal
processing for digital communications, and in particular: xDSL systems,
resource allocation, synchronization/estimation, and multicarrier
modulations.
Prof. Louveaux was a co-recipient of the "Prix biennal Siemens 2000" for
a contribution on filter-bank based multi-carrier transmission and
co-recipient of the the "Prix Scientifique Alcatel 2005" for a
contribution in the field of powerline communications.
\end{IEEEbiography}

\begin{IEEEbiography}[{\includegraphics[width=1in,height=1.5in,clip,keepaspectratio]{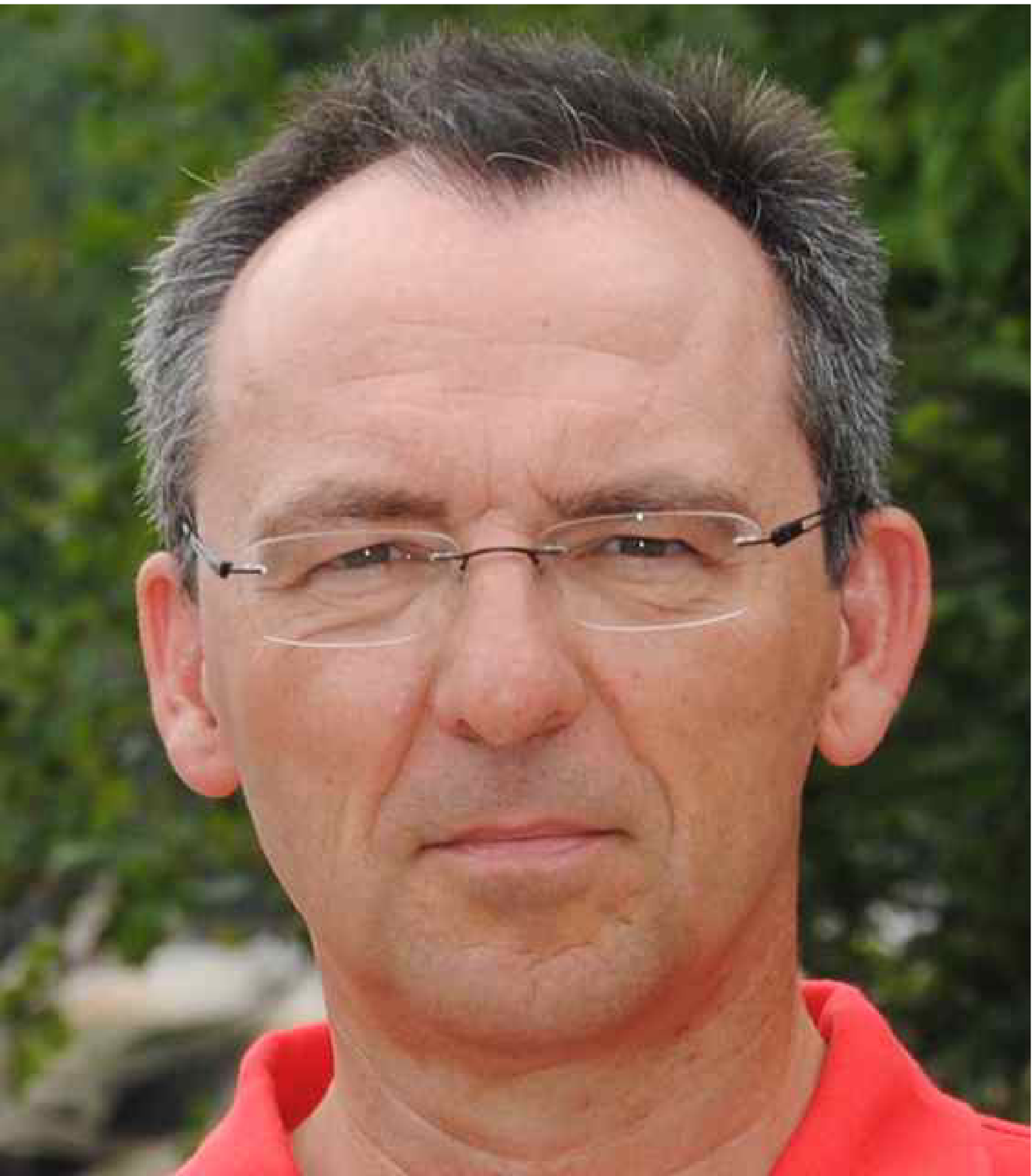}}]{Luc Vandendorpe}
was born in Mouscron, Belgium in 1962. He received the Electrical Engineering degree ({\it summa cum laude})
and the Ph. D. degree from the Universit?Catholique de Louvain (UCL) Louvain-la-Neuve, Belgium
in 1985 and 1991 respectively. Since 1985, he is with the Communications and Remote
Sensing Laboratory of UCL where he first worked in the field of bit rate reduction techniques
for video coding. In 1992, he was a Visiting Scientist and Research Fellow at the Telecommunications
and Traffic Control Systems Group of the Delft Technical University, The Netherlands, where he worked
on Spread Spectrum Techniques for Personal Communications Systems.
From October 1992 to August 1997, L. Vandendorpe was Senior Research Associate of the Belgian
NSF at UCL, and invited assistant professor. Presently he is Professor and head of
the Institute for Information and Communication Technologies, Electronics and Applied Mathematics.

His current interest is in digital communication systems and more precisely resource
allocation for OFDM(A) based multicell systems, MIMO and distributed MIMO,
sensor networks, turbo-based communications systems, physical layer security and UWB based positioning.
In 1990, he was co-recipient of the Biennal Alcatel-Bell Award from the Belgian NSF
for a contribution in the field of image coding. In 2000 he was co-recipient
(with J. Louveaux and F. Deryck) of the Biennal Siemens Award from the Belgian NSF
for a contribution about filter bank based multicarrier transmission.
In 2004 he was co-winner (with J. Czyz) of the Face Authentication Competition,
FAC 2004. L. Vandendorpe is or has been TPC member for numerous IEEE conferences
(VTC Fall, Globecom Communications Theory Symposium, SPAWC, ICC) and for the Turbo Symposium.
He was co-technical chair (with P. Duhamel) for IEEE ICASSP 2006.
He was an editor of the IEEE Trans. on Communications for Synchronisation and
Equalization between 2000 and 2002, associate editor of the IEEE Trans. on
Wireless Communications between 2003 and 2005, and associate editor of the
IEEE Trans. on Signal Processing between 2004 and 2006. He was chair of the
IEEE Benelux joint chapter on Communications and Vehicular Technology
 between 1999 and 2003. He was an elected member of the Signal Processing for
Communications committee between 2000 and 2005, and an elected member of the Sensor
Array and Multichannel Signal Processing committee of the Signal Processing Society between 2006 and 2008.
He was an elected member of the Signal Processing for
Communications committee between 2000 and 2005,
and between 2009 and 2011, and an elected member
of the Sensor Array and Multichannel Signal Processing committee of the
Signal Processing Society between 2006 and 2008. He is the
Editor in Chief for the EURASIP Journal on Wireless Communications and
Networking and a Fellow of the IEEE.
\end{IEEEbiography}

% that's all folks
\end{document}